\documentclass[iop]{emulateapj}

\usepackage{graphicx}

\newcommand{\msunyr}{\ensuremath{\mathit{M}_{\odot}{\rm \ yr}^{-1}}}   
\newcommand{\kms}{\ensuremath{{\rm km \ s^{-1}}}}                   
\newcommand{\msun}{\ensuremath{\mathit{M}_{\odot}}}               
\newcommand{\lsun}{\ensuremath{\mathit{L}_{\odot}}}                  

\newcommand{\lstar}{\ensuremath{\mathit{L}_{\star}}}                 
\newcommand{\mdot}{\ensuremath{\dot{M}}}                             
\newcommand{\teff}{\ensuremath{\mathit{T}_{\rm eff}}}                
\newcommand{\vinf}{\ensuremath{v_{\infty}}}                          
\newcommand{\tstar}{\ensuremath{\mathit{T}_{\star}}}                 

\newcommand{\vrot}{\ensuremath{v_{\rm rot}}}                         
\newcommand{\vcrit}{\ensuremath{v_{\rm crit}}}                         

\shorttitle{Is Eta Car a fast rotator?}
\shortauthors{Groh et al.}
\slugcomment{Draft version May 21, 2010}
\begin{document}

\title{Is Eta Carinae a fast rotator, and how much does the companion influence the inner wind structure?\altaffilmark{1}}

\author{J. H. Groh\altaffilmark{2}, T. I. Madura\altaffilmark{3}, S. P. Owocki\altaffilmark{3}, D. J. Hillier\altaffilmark{4}, and G. Weigelt\altaffilmark{2}}

\altaffiltext{1}{Based on observations made with VLTI/VINCI and VLTI/AMBER.}
\altaffiltext{2}{Max-Planck-Institut fuer Radioastronomie, Auf dem Huegel 69, D-53121 Bonn, Germany; email: jgroh@mpifr.de}
\altaffiltext{3}{Bartol Research Institute, University of Delaware, Newark, DE 19716, USA}
\altaffiltext{4}{Department of Physics and Astronomy, University of Pittsburgh, 3941 O'Hara Street, Pittsburgh, PA 15260, USA}

\begin{abstract}
We analyze interferometric measurements of the Luminous Blue Variable Eta Carinae with the goal of constraining the rotational velocity of the primary star and probing the influence of the companion. Using 2-D radiative transfer models of latitude-dependent stellar winds, we find that prolate wind models with a ratio of the rotational velocity ($\vrot$) to the critical velocity ($\vcrit$) of $W=0.77-0.92$, inclination angle of $i=60\degr-90\degr$, and position angle PA $= 108\degr-142\degr$ reproduce simultaneously K-band continuum visibilities from VLTI/VINCI and closure phase measurements from VLTI/AMBER. Interestingly, oblate models with $W=0.73-0.90$ and $i=80\degr-90\degr$ produce similar fits to the interferometric data, but require PA $= 210\degr-230\degr$. Therefore, both prolate and oblate models suggest that the rotation axis of the primary star is not aligned with the Homunculus polar axis. We also compute radiative transfer models of the primary star allowing for the presence of a cavity and dense wind-wind interaction region created by the companion star. We find that the wind-wind interaction has a significant effect on the $K$-band image mainly via free-free emission from the compressed walls and, for reasonable model parameters, can reproduce the VLTI/VINCI visibilities taken at $\phi_\mathrm{vb03}=0.92-0.93$. We conclude that the density structure of the primary wind can be sufficiently disturbed by the companion, thus mimicking the effects of fast rotation in the interferometric observables. Therefore, fast rotation may not be the only explanation for the interferometric observations. Intense temporal monitoring and 3-D modeling are needed to resolve these issues.
\end{abstract}


\section{Introduction} 

\defcitealias{bh05}{BH05}
\defcitealias{hillier06}{H06}
\defcitealias{hillier01}{H01}
\defcitealias{vb03}{VB03}
\defcitealias{weigelt07}{W07}
\defcitealias{smith03}{S03}

Eta Carinae is one of the most luminous objects in the Galaxy, allowing for the study of massive stellar evolution under extreme conditions. Eta Car also presents us with a unique opportunity to witness the early evolution of violent giant outbursts that happen during the Luminous Blue Variable (LBV) phase, like that which created Eta Car's massive Homunculus nebula \citep{thackeray50,smith03b}. Eta Car is generally accepted to be a binary system \citep{damineli97} comprised of two massive stars, $\eta_\mathrm{A}$ (primary) and $\eta_\mathrm{B}$ (secondary). The total luminosity is dominated by $\eta_\mathrm{A}$ and amounts to $L_{\mathrm{tot}}\geq 5\times 10^6~\lsun$ \citep{dh97}, and the total mass is of at least $110~\msun$ \citep[][hereafter H01]{hillier01}. Most authors agree on a high eccentricity ($e\sim0.9$, \citealt{corcoran01}) and an orbital period of $2022.7 \pm 1.3$ days \citep{damineli08_period}, while the other orbital parameters are uncertain.

$\eta_\mathrm{A}$ is an LBV star with a high mass-loss rate of $\mdot \simeq 10^{-3} \msunyr$ and wind terminal velocity of $\vinf \simeq 500-600 \ \kms$ \citepalias{hillier01}. The stellar parameters and evolutionary state of $\eta_\mathrm{B}$ are uncertain since it has never been directly detected, and only indirect information is available. A broad range of $\eta_\mathrm{B}$ stellar parameters can explain the ionization of the inner ejecta \citep{mehner10}, and X-ray observations require $\eta_\mathrm{B}$ to have $\vinf\simeq 3000~\kms$ and $\mdot \simeq 10^{-5}~\msunyr$ \citep[e.g.,][]{pc02}. 3-D numerical simulations show that the wind of $\eta_\mathrm{B}$ influences the geometry of the very dense wind of $\eta_\mathrm{A}$ via the creation of a cavity and dense wind-wind collision zone \citep{pc02,okazaki08,parkin09}.

Based on the variations of H$\alpha$ absorption line profiles in scattered light from the Homunculus, which provide us with different viewing directions to the star, and assuming that the rotation axis of $\eta_\mathrm{A}$ and the Homunculus axis are aligned, \citet[][hereafter S03]{smith03} suggested that $\eta_\mathrm{A}$ has a latitude-dependent wind, with faster, denser outflow in polar directions. Interferometric measurements obtained with VLTI/VINCI (\citealt{vb03}, hereafter VB03; \citealt{kervella07}) and VLTI/AMBER \citep[][hereafter W07]{weigelt07} in the near-infrared $K$-band continuum are consistent with an ellipsoidal shape projected on the sky. Both \citetalias{vb03} and \citetalias{weigelt07} interpreted this as evidence for a dense prolate wind generated by fast rotation, as theoretically predicted by \citet{owocki96,owocki98}. However, the influence of $\eta_\mathrm{B}$ on the H$\alpha$ absorption profiles and on the $K$-band emission, which are the key diagnostics supporting the fast rotation of $\eta_\mathrm{A}$, is poorly constrained.

We analyze published K-band continuum interferometric data of Eta Car obtained at orbital phases\footnote{Assuming the ephemeris from \citet{damineli08_period}.} $\phi_\mathrm{vb03}=0.92$--0.93 \citepalias{vb03} and $\phi_\mathrm{w07}=0.27$--0.30 \citepalias{weigelt07}. Our goals are to constrain the rotational velocity and spatial orientation of the rotation axis of $\eta_\mathrm{A}$ based on the effects of rotation on the wind density structure, which determines the geometry of the $K$-band emitting region. We also investigate the influence of $\eta_\mathrm{B}$ on the inner wind of $\eta_\mathrm{A}$ through the presence of a low-density cavity and density enhanced wind-wind collision zone, with the goal of determining how these may affect the interpretation of the VINCI dataset obtained relatively close to periastron. We apply, for the first time for Eta Car, 2-D radiative transfer models of latitude-dependent winds generated by rapid rotation (Section \ref{radtranslat}) and of a modified wind of $\eta_\mathrm{A}$ which includes a wind cavity and colliding wind interaction region (Section \ref{radtranscav}). Sections \ref{reslat} and \ref{rescav} compare the computed interferometric observables for each scenario with the available observations. Section \ref{disc} discusses whether $\eta_\mathrm{A}$ is a fast rotator, a possible misalignment between its rotation axis and the Homunculus polar axis, and the influence of $\eta_\mathrm{B}$ on the wind of $\eta_\mathrm{A}$.

\section{Single-star, latitude-dependent wind generated by rapid rotation} \label{latidep}

\subsection{Radiative transfer modeling} \label{radtranslat}

We compute 2-D latitude-dependent wind models for $\eta_\mathrm{A}$ using an updated version of the 2-D radiative transfer code of \citet[hereafter BH05]{bh05} and via a methodology similar to that described in \citet{groh08}. In the following, we briefly describe only the essential aspects of the code.  We refer the reader to \citetalias{bh05},  \citet{ghd06,groh08,gdh09},  and \citet{driebe09} for further details on the code's applications.

The 2-D models use as input several quantities (e.g., energy-level populations, ionization structure, and radiation field) from the spherically symmetric model of $\eta_\mathrm{A}$ \citepalias{hillier01,hillier06} computed using the non-LTE, fully line blanketed radiative transfer code {\sc CMFGEN} \citep{hm98}. We assume the same parameters derived by \citetalias{hillier01}:  stellar temperature \tstar~=~35,310~K (at Rosseland optical depth $\tau_\mathrm{Ross}=150$), effective temperature \teff = 9,210~K (at $\tau_\mathrm{Ross} = 2/3$), luminosity $\lstar = 5 \times 10^6~\lsun$, $\mdot = 10^{-3}~\msunyr$, $\vinf = 500~\kms$, a clumping volume-filling factor $f = 0.1$, and distance $d = 2.3$ kpc. The atomic model and abundances are described by \citetalias{hillier01} and \citetalias{hillier06}.

The code allows for the specification of any arbitrary latitude-dependent variation of the wind density $\rho$ and $\vinf$. For the latitudinal variation of $\rho$, we adopt the predictions from \citet{owocki98} for gravity-darkened (GD) line-driven prolate winds,
\begin{equation}
\mathrm{GD,~prolate}: \,\,\, \frac{\rho(\theta)}{\rho_{0}} \propto  \sqrt{1 - W^2 \sin^{2}\theta}\,\,,
\end{equation}
and the non-gravity-darkened (non-GD) scaling for oblate winds,
\begin{equation}
\mathrm{non-GD,~oblate}: \,\,\, \frac{\rho(\theta)}{\rho_{0}} \propto \frac{1}{1 - W^2 \sin^{2}\theta}\,\,,
\end{equation}
where $\theta$ is the colatitude angle (0$^{\circ}$= pole, 90$^{\circ}$= equator), $\rho_0$ is the density at the pole, and $W$ is the ratio of the rotational velocity of $\eta_\mathrm{A}$ ($\vrot$) to its critical velocity ($\vcrit$). Note that the oblate-wind density scaling assumes a standard CAK \citep{CAK} $\alpha$ parameter value of $2/3$. The inclination angle $i$ is the tilt angle between the rotation axis of the star and the line of sight ($i = 0^{\circ}$= pole-on view, $i = 90^{\circ}$= equator-on view). Emissivities and opacities are scaled according to the modified density, {but the code assumes a spherically symmetric ionization structure.} Gravity darkening and distortion of the shape of the hydrostatic core of the star are not included at this time; however, these are not expected to significantly affect the $K$-band emitting region since the source function in the infrared continuum depends only weakly on the temperature. For consistency, we also assume the \citet{owocki98} predictions for $\vinf(\theta)$.

We calculated a grid of prolate and oblate models with $i$ ranging from $0\degr$ to $90\degr$ in steps of $1\degr$, and $W$ ranging from 0 to 0.99 in steps of 0.01, totaling 18,200 models. For a desired position angle (PA) orientation on the sky\footnote{Measured in degrees east of north.}, the $K$-band image, visibilities, and closure phases (CP) are computed.

\begin{figure*}
\centering
\resizebox{0.245\hsize}{!}{\includegraphics{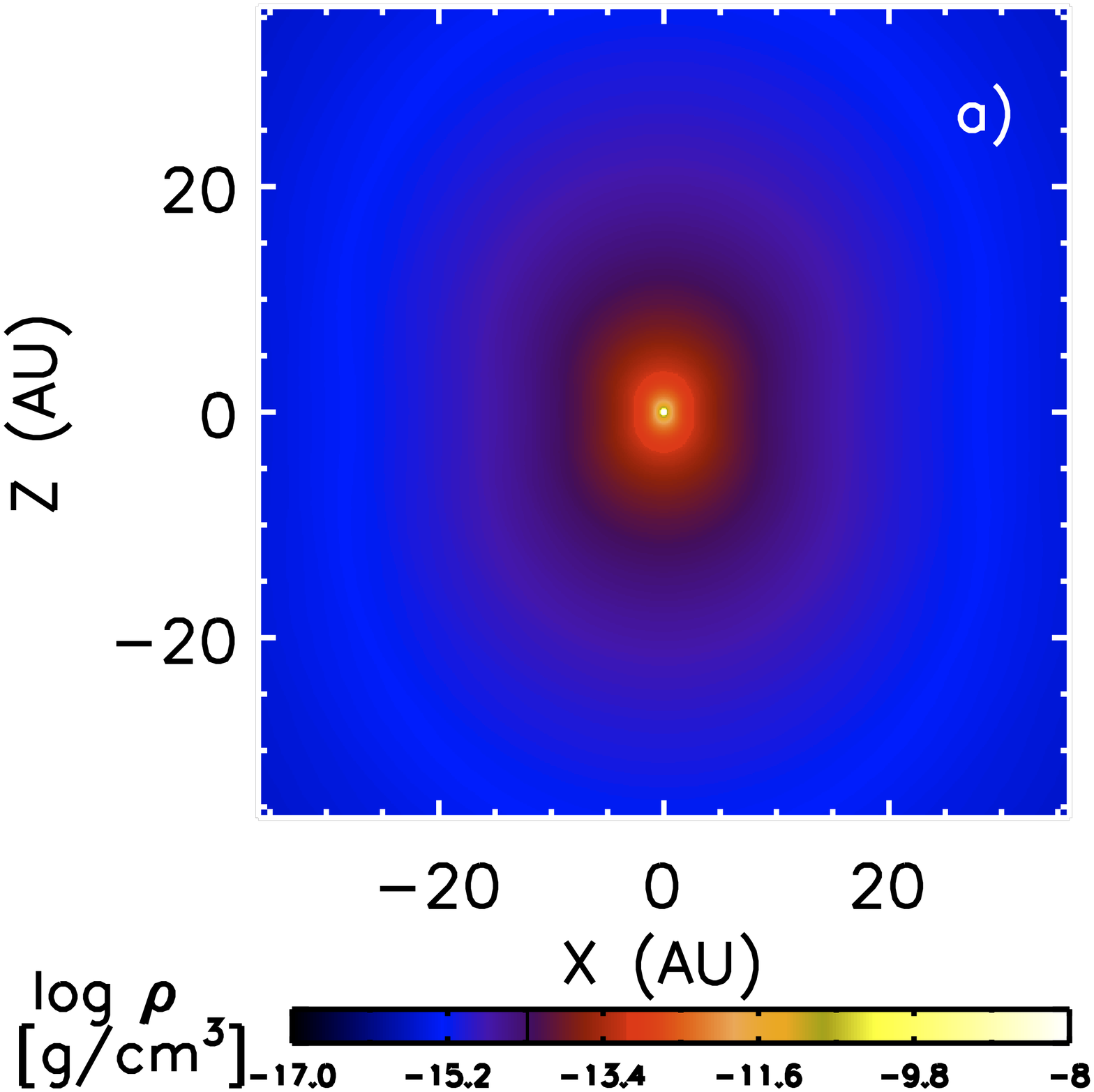}}
\resizebox{0.245\hsize}{!}{\includegraphics{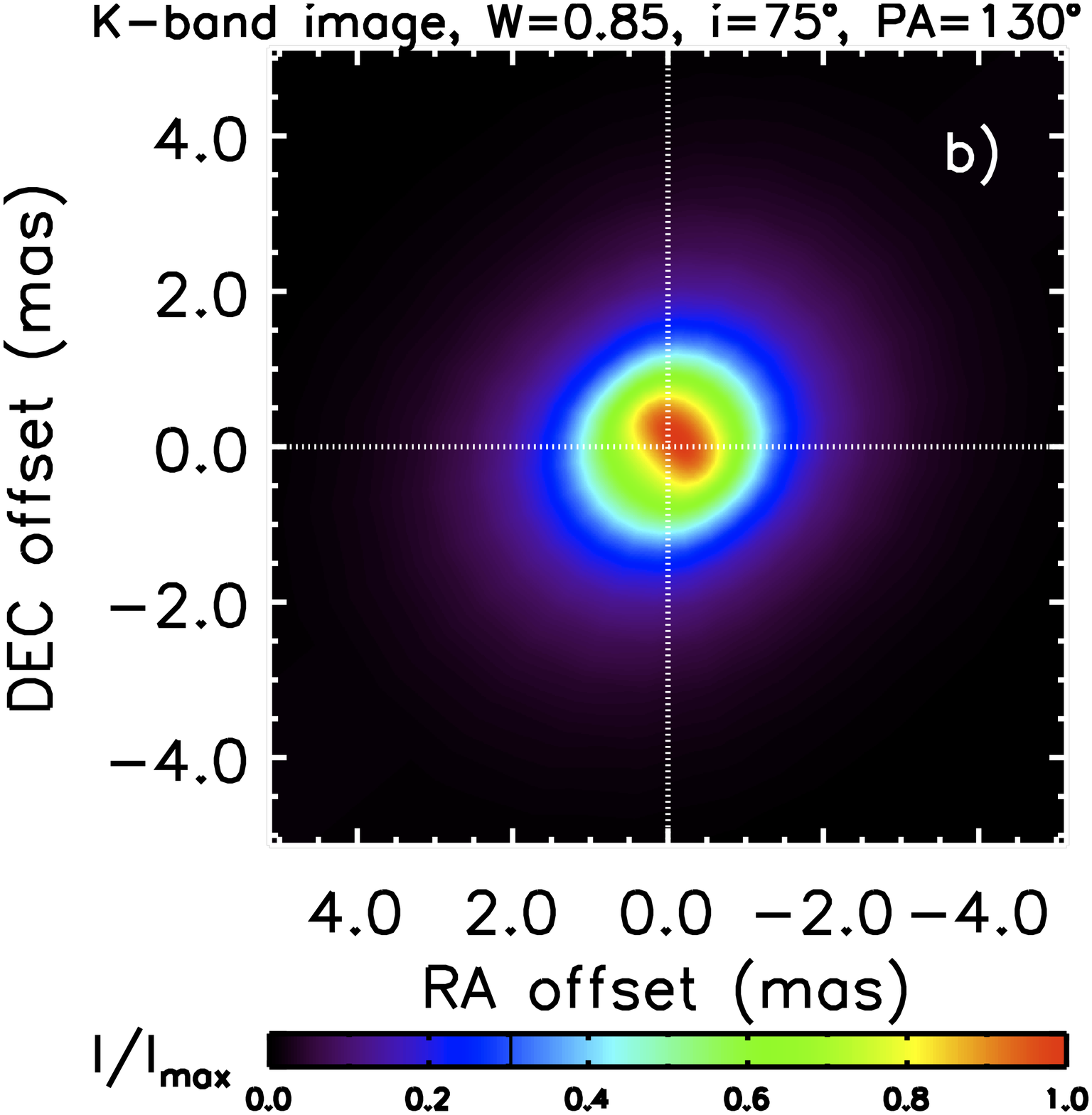}}
\resizebox{0.245\hsize}{!}{\includegraphics{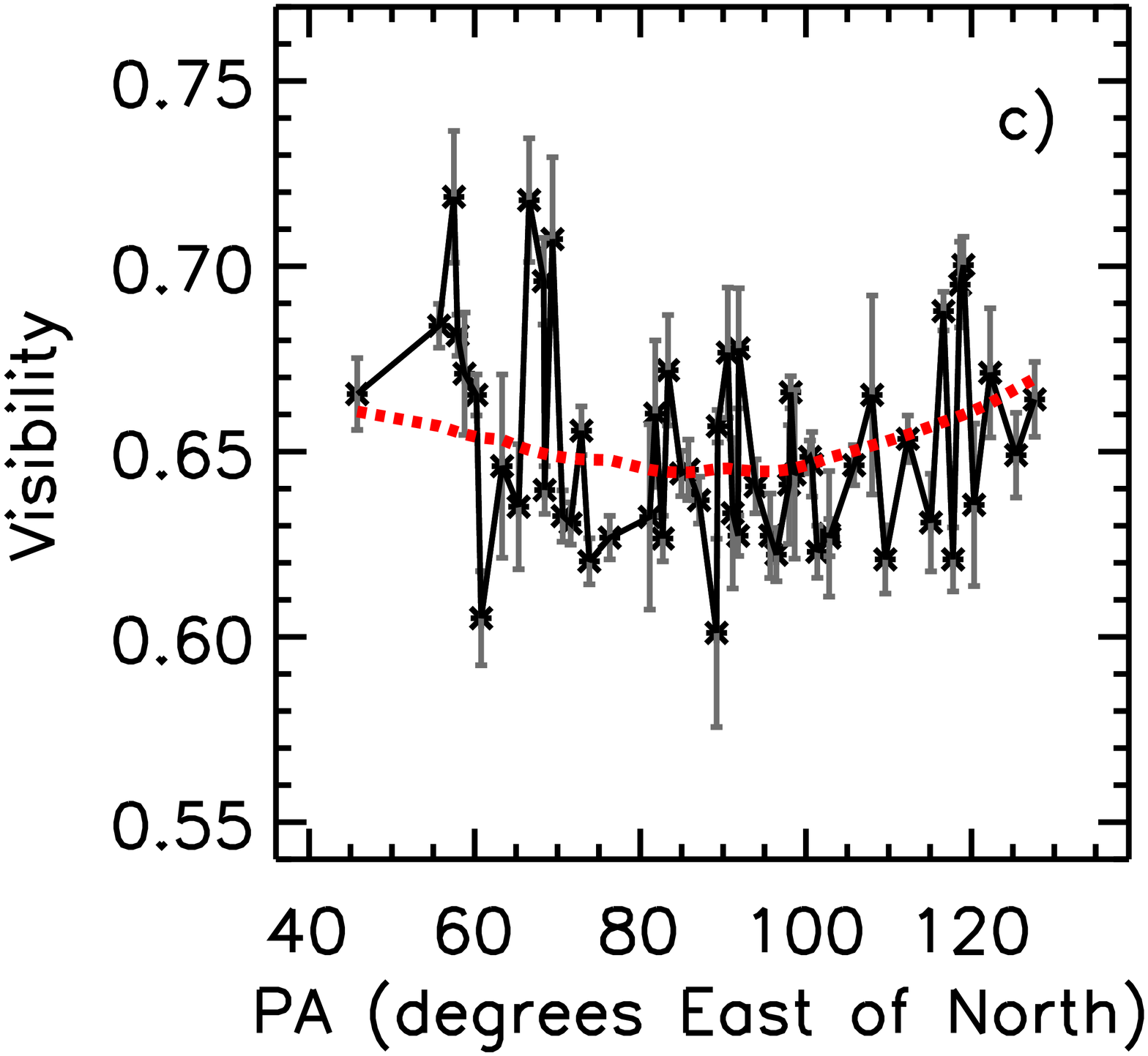}}\\
\resizebox{0.245\hsize}{!}{\includegraphics{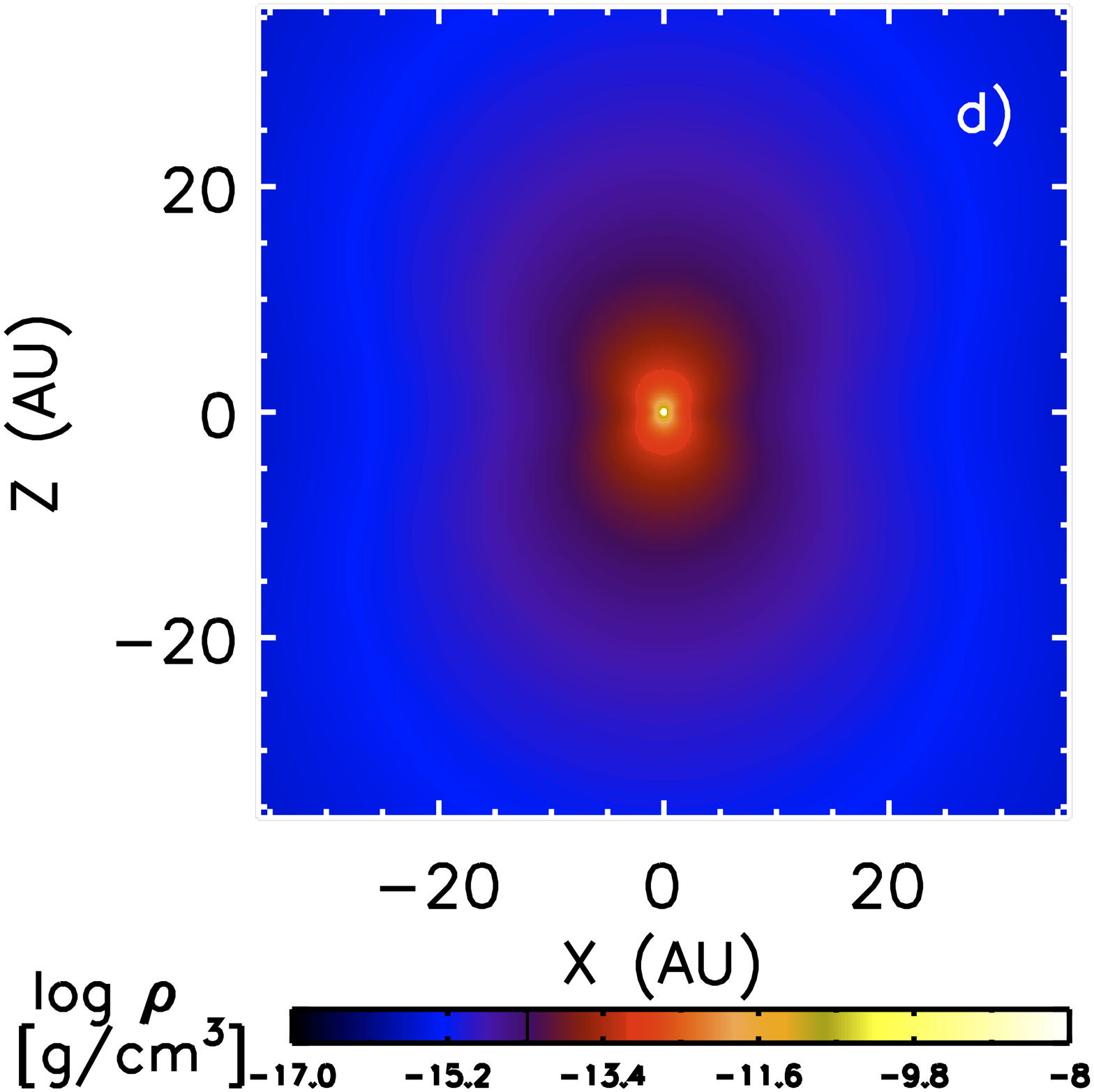}}
\resizebox{0.245\hsize}{!}{\includegraphics{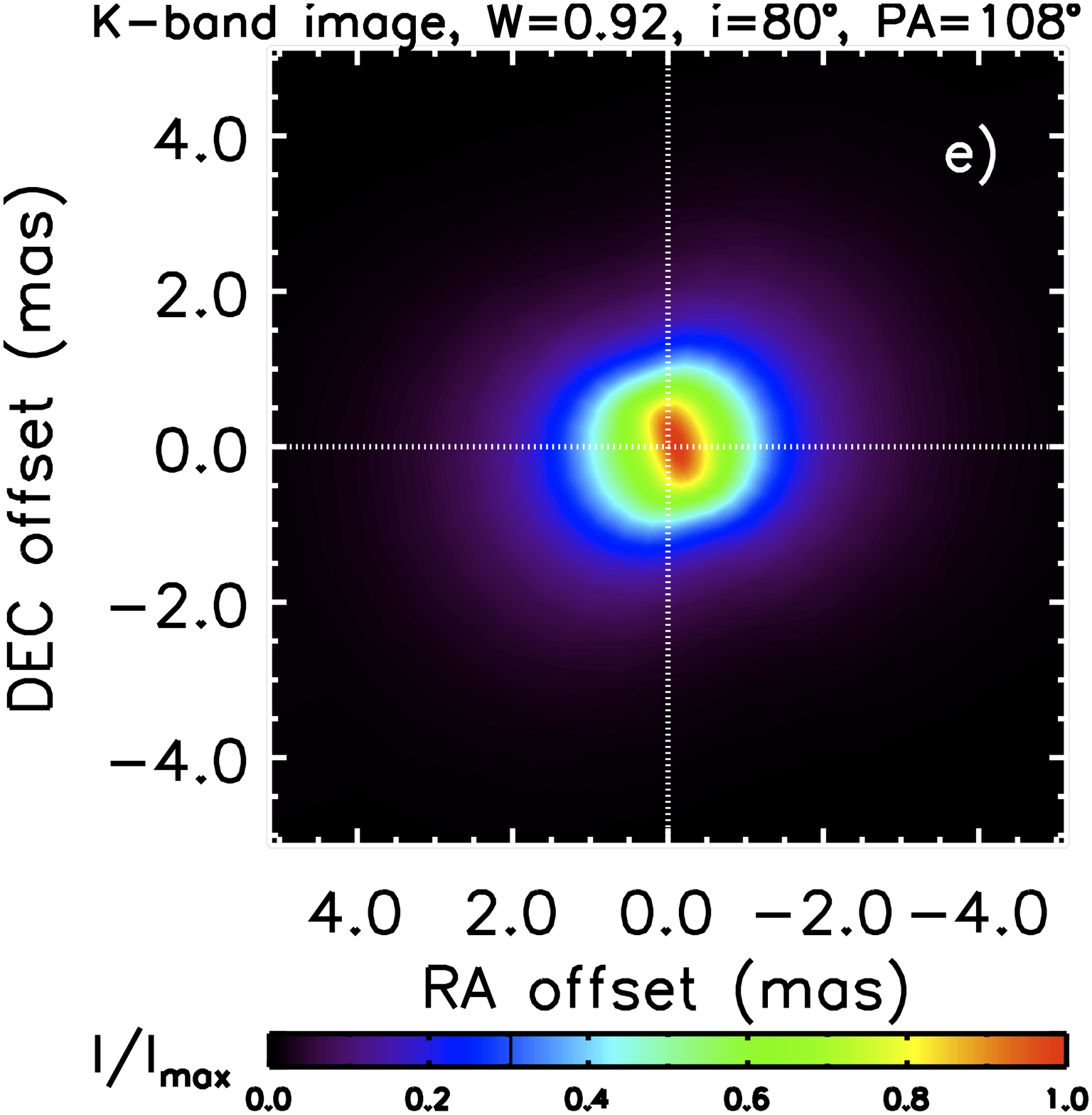}}
\resizebox{0.245\hsize}{!}{\includegraphics{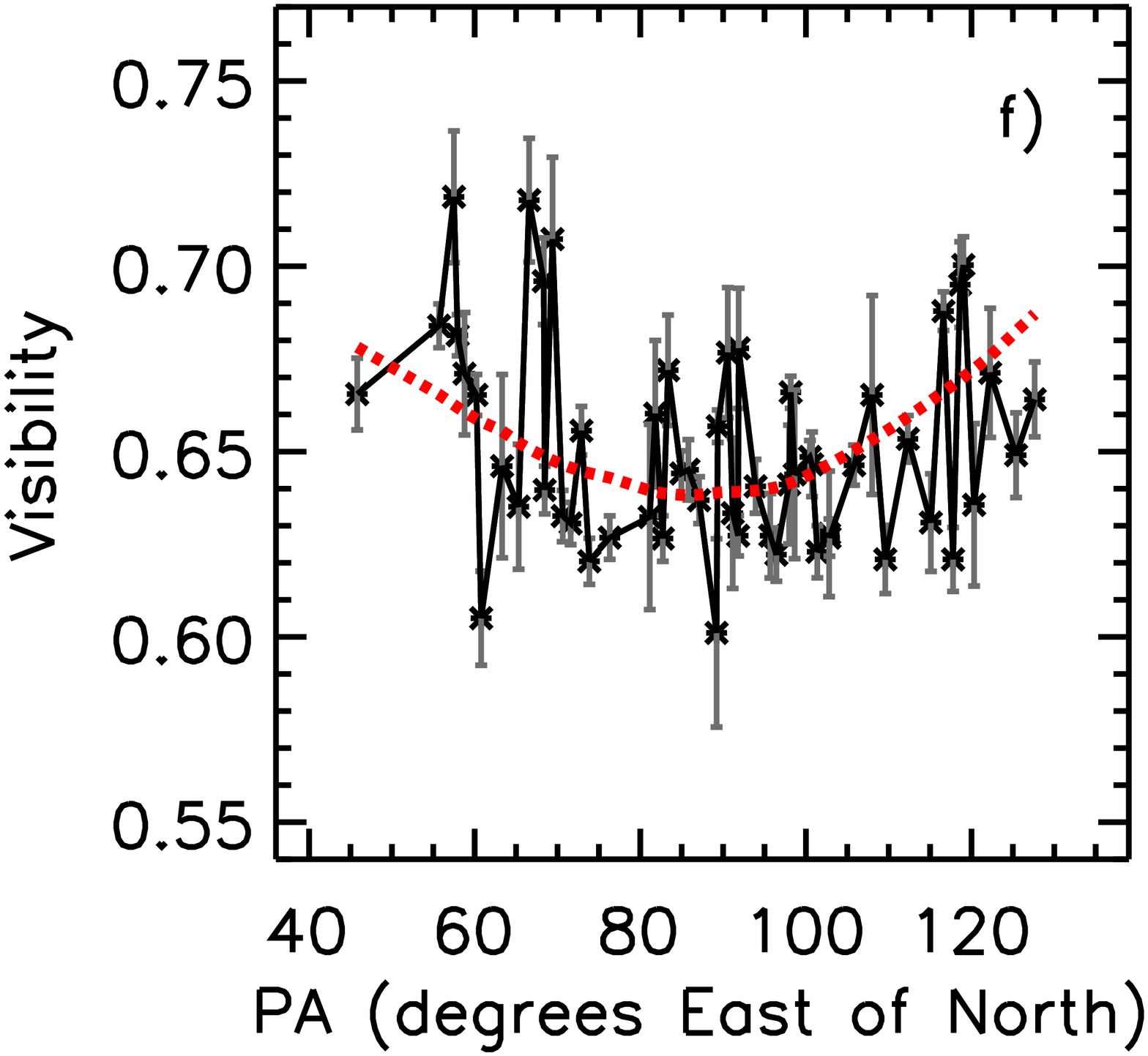}}\\
\resizebox{0.245\hsize}{!}{\includegraphics{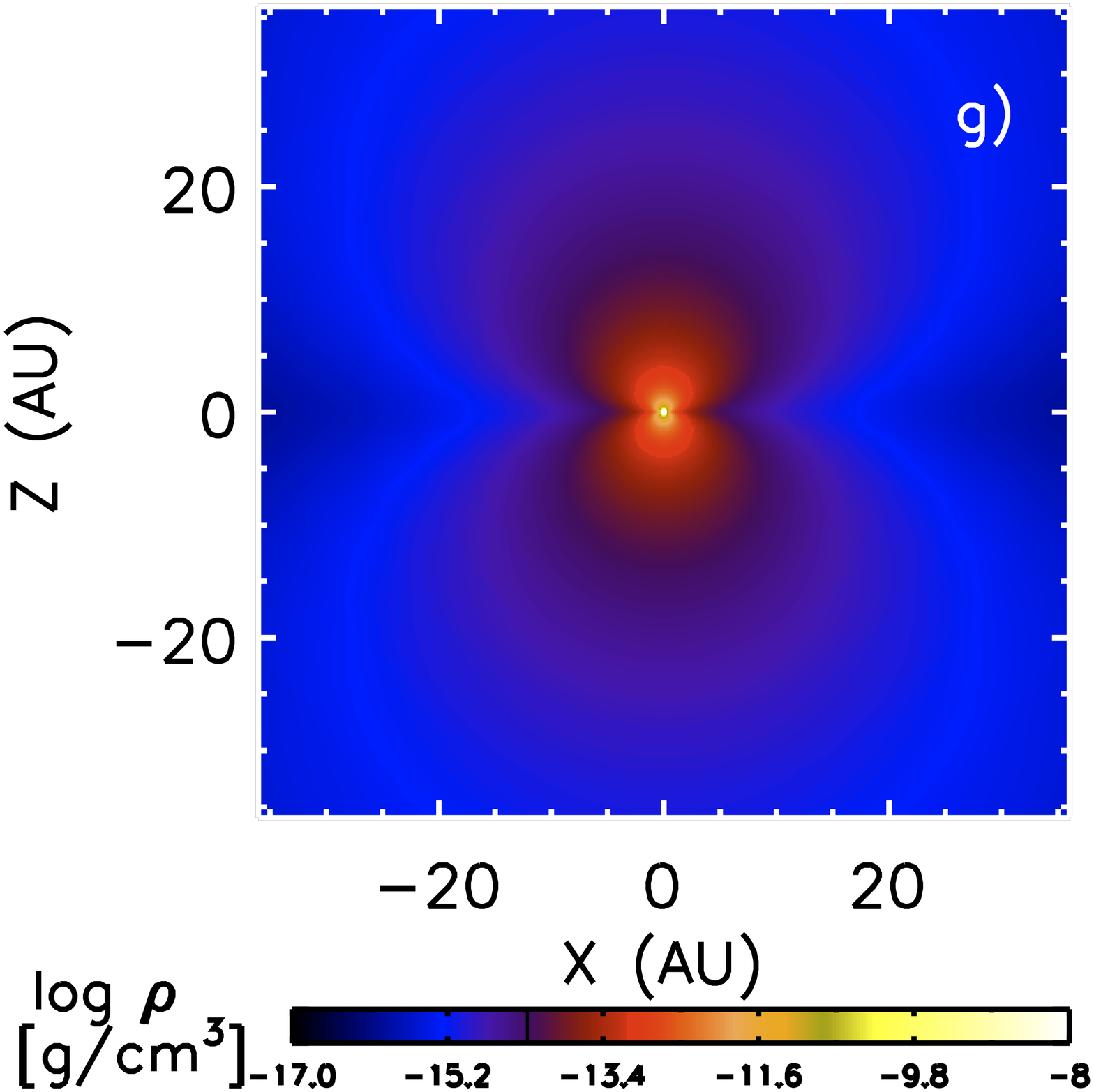}}
\resizebox{0.245\hsize}{!}{\includegraphics{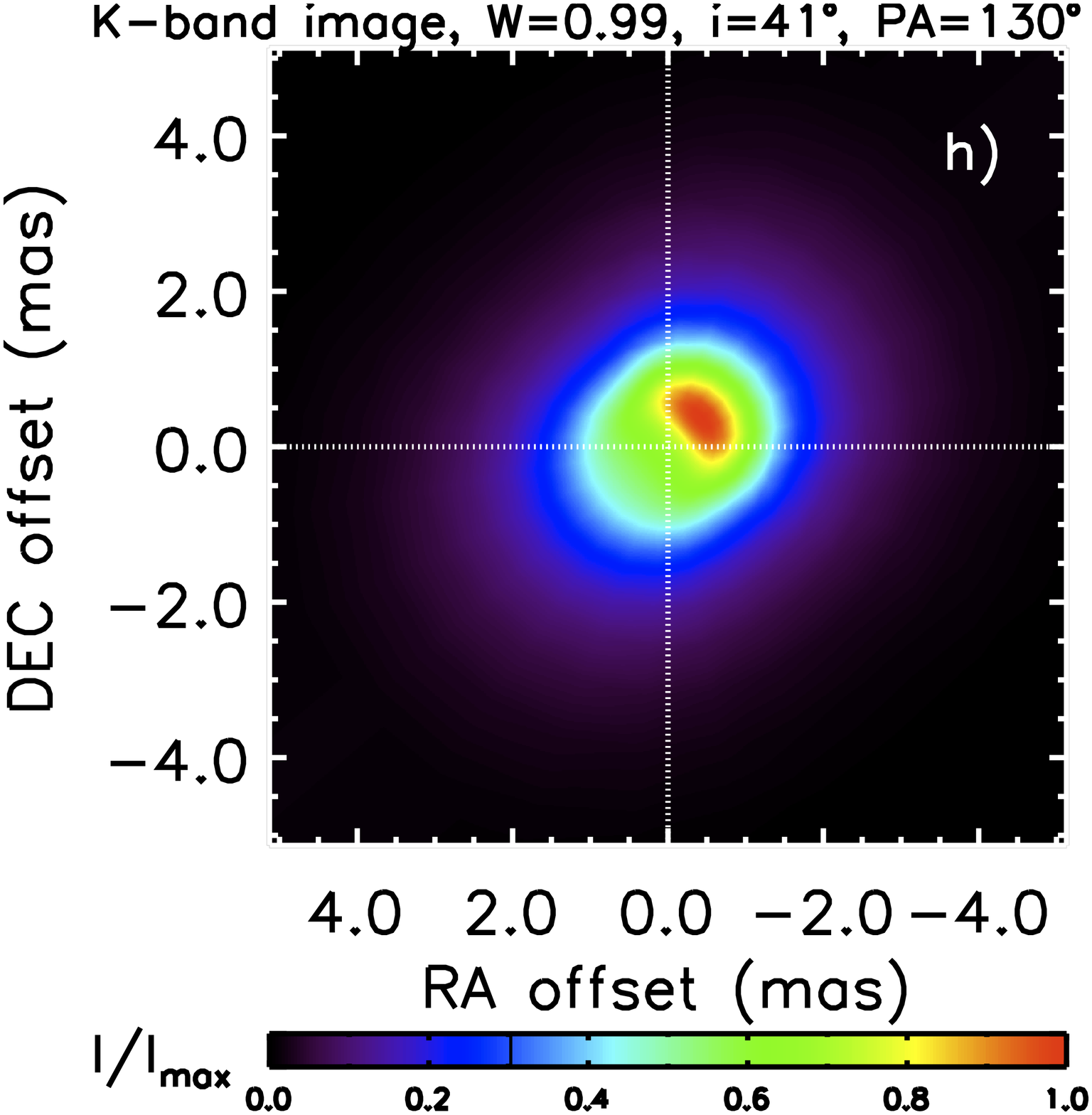}}
\resizebox{0.245\hsize}{!}{\includegraphics{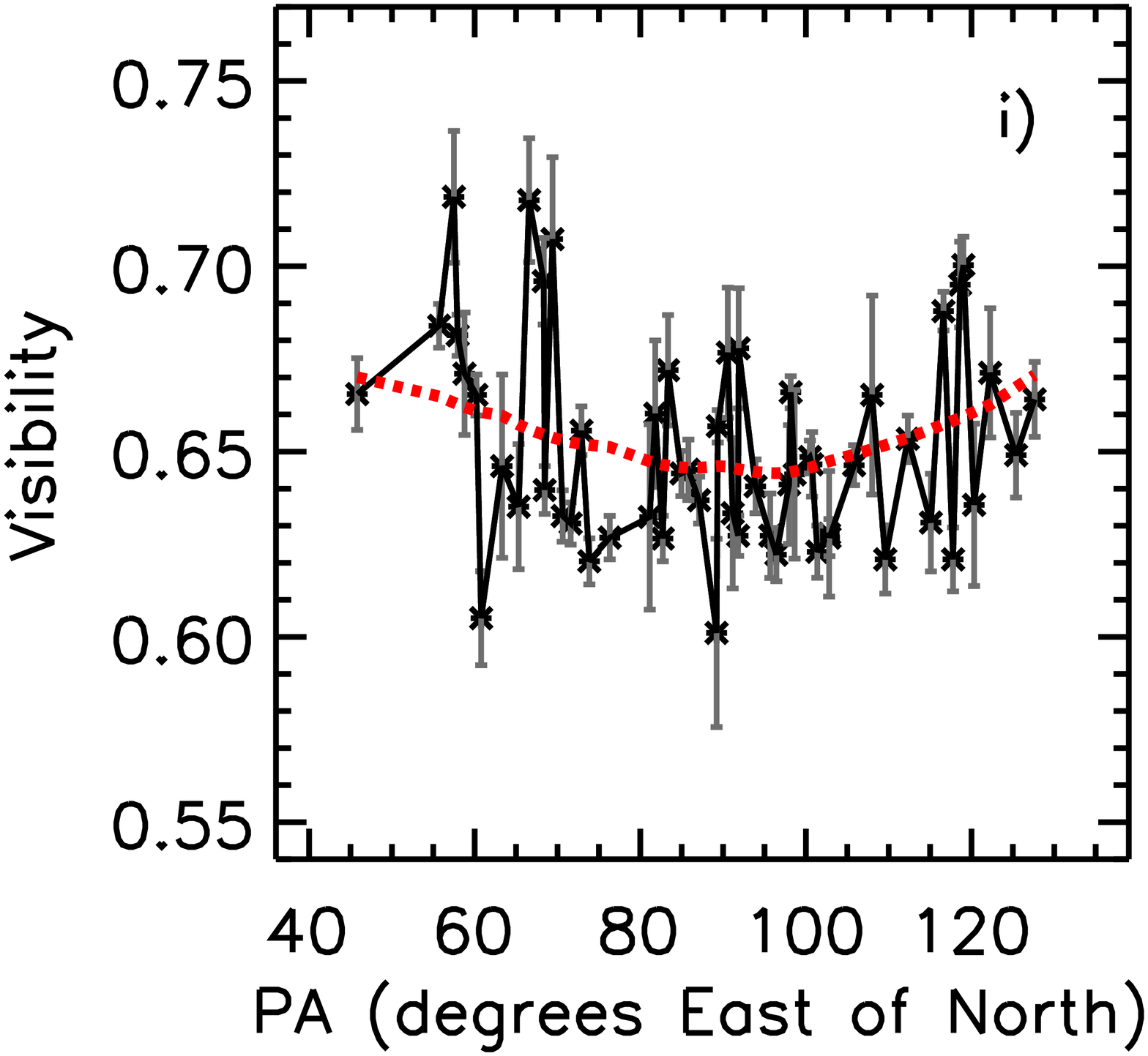}}\\
\caption{\label{figlatiden}{\it Left panels:} Density structure of the latitude-dependent wind models of $\eta_\mathrm{A}$ in the x-z plane (i.\,e., equator-on). {\it Middle:} K-band image projected on the sky. {\it Right:} VINCI visibilities for the 24~m baseline as a function of telescope PA (connected black  asterisks) compared to the respective model prediction (red dotted line). Note that the projected baseline length of the VINCI measurements changes as a function of PA. From top to bottom, the panels correspond to examples of best-fit prolate wind models with $W = 0.85 $, $i=75\degr$, $\mathrm{PA}=130\degr$ (panels $a-c$) and $W = 0.92 $, $i=80\degr$, $\mathrm{PA}=108\degr$ (panels $d-f$), and prolate wind aligned with the Homunculus but with large closure phase (panels $g-i$; $W = 0.99 $, $i=41\degr$, $\mathrm{PA}=130\degr$).}
\end{figure*}

\subsection{Results}\label{reslat}

\begin{figure}
\centering
\resizebox{0.485\hsize}{!}{\includegraphics{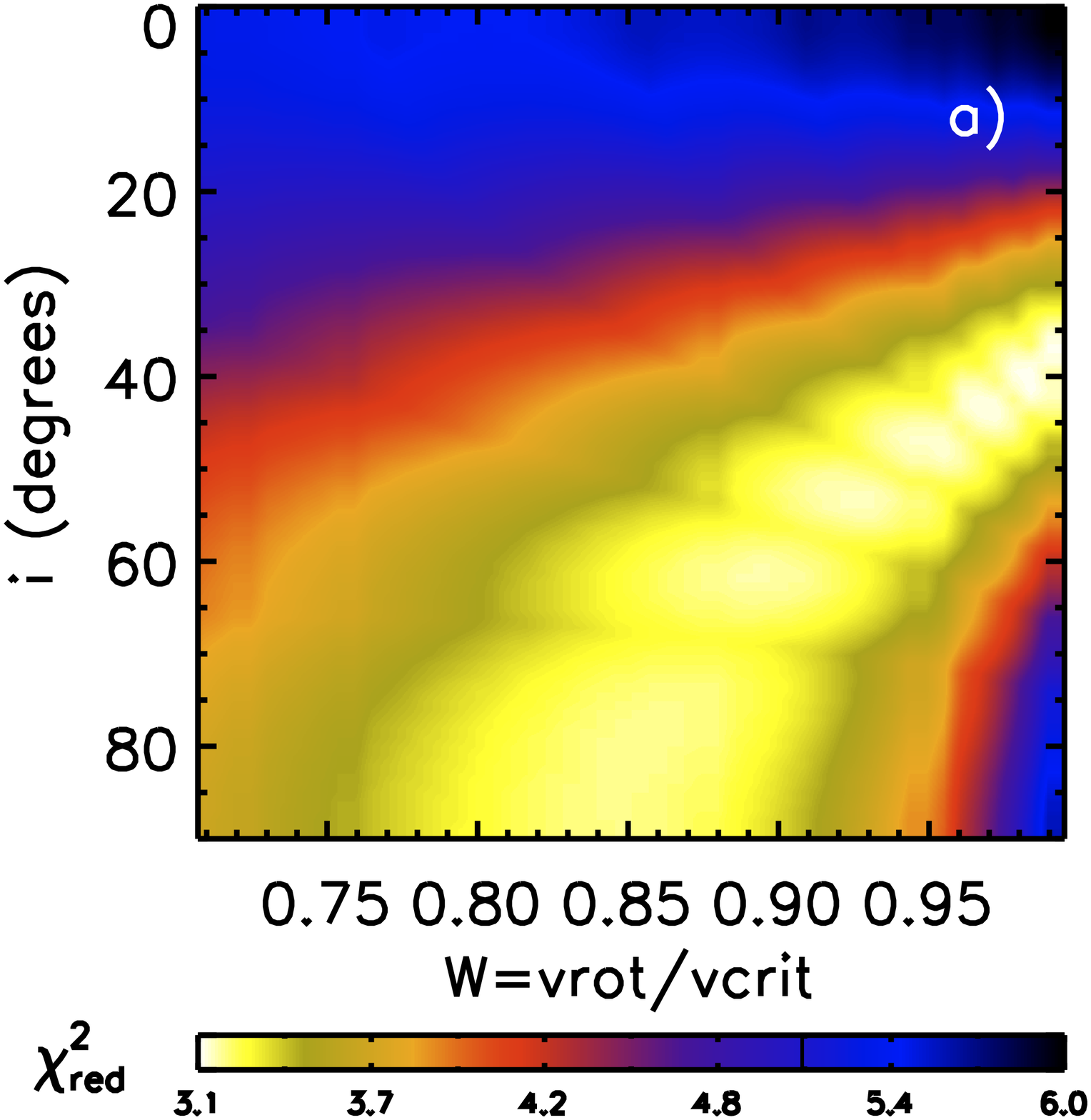}}
\resizebox{0.485\hsize}{!}{\includegraphics{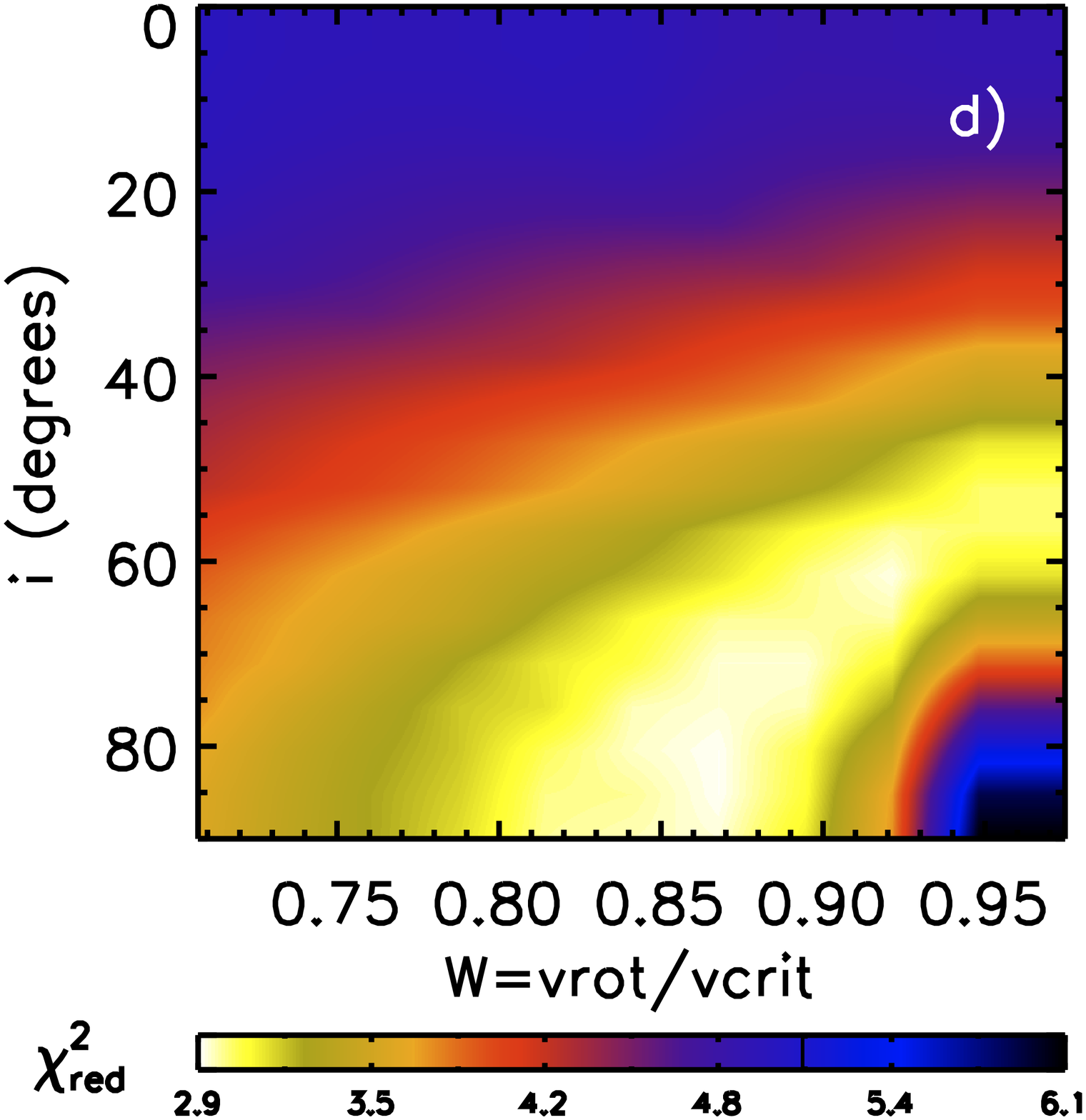}}\\
\resizebox{0.485\hsize}{!}{\includegraphics{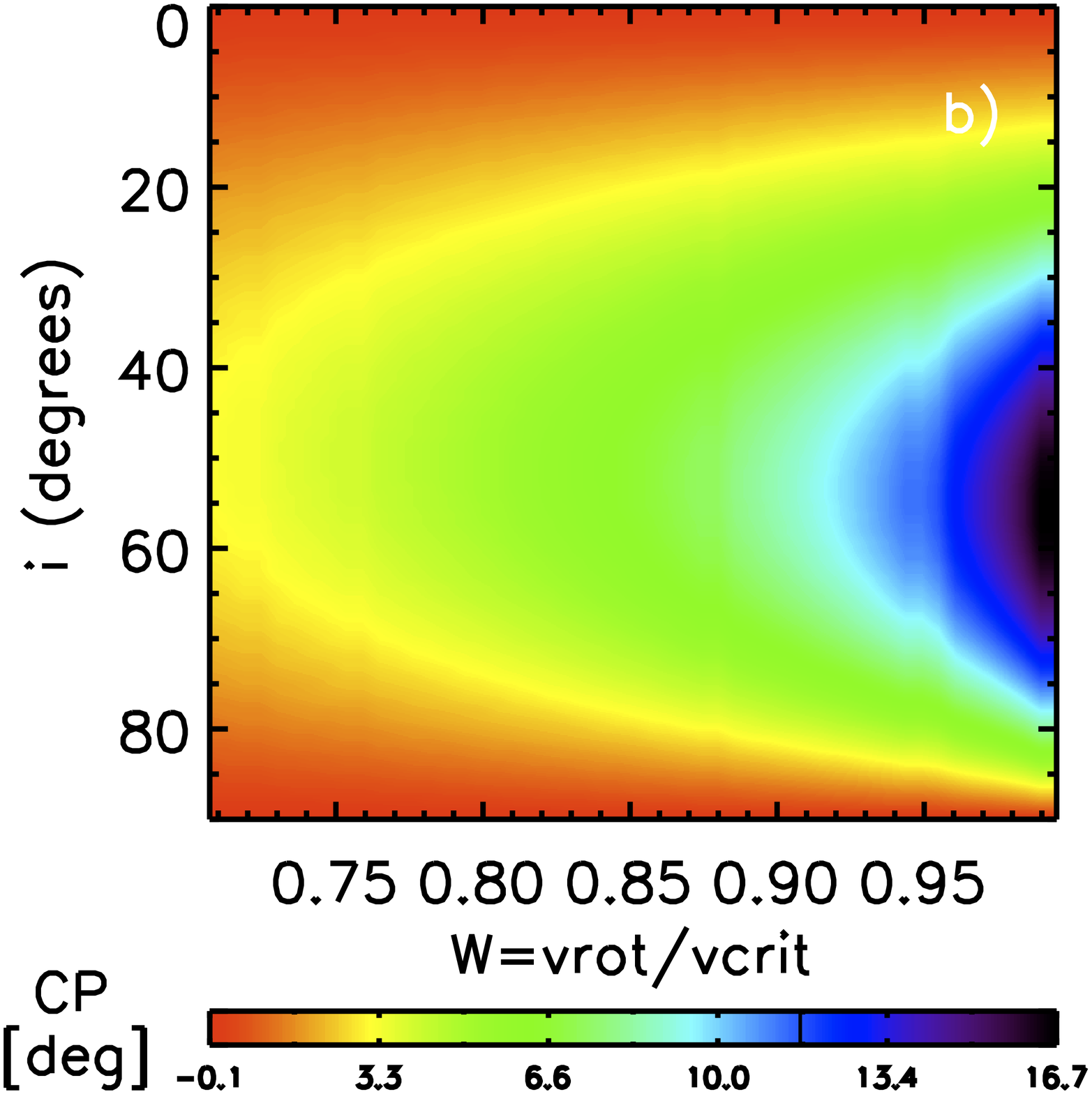}}
\resizebox{0.485\hsize}{!}{\includegraphics{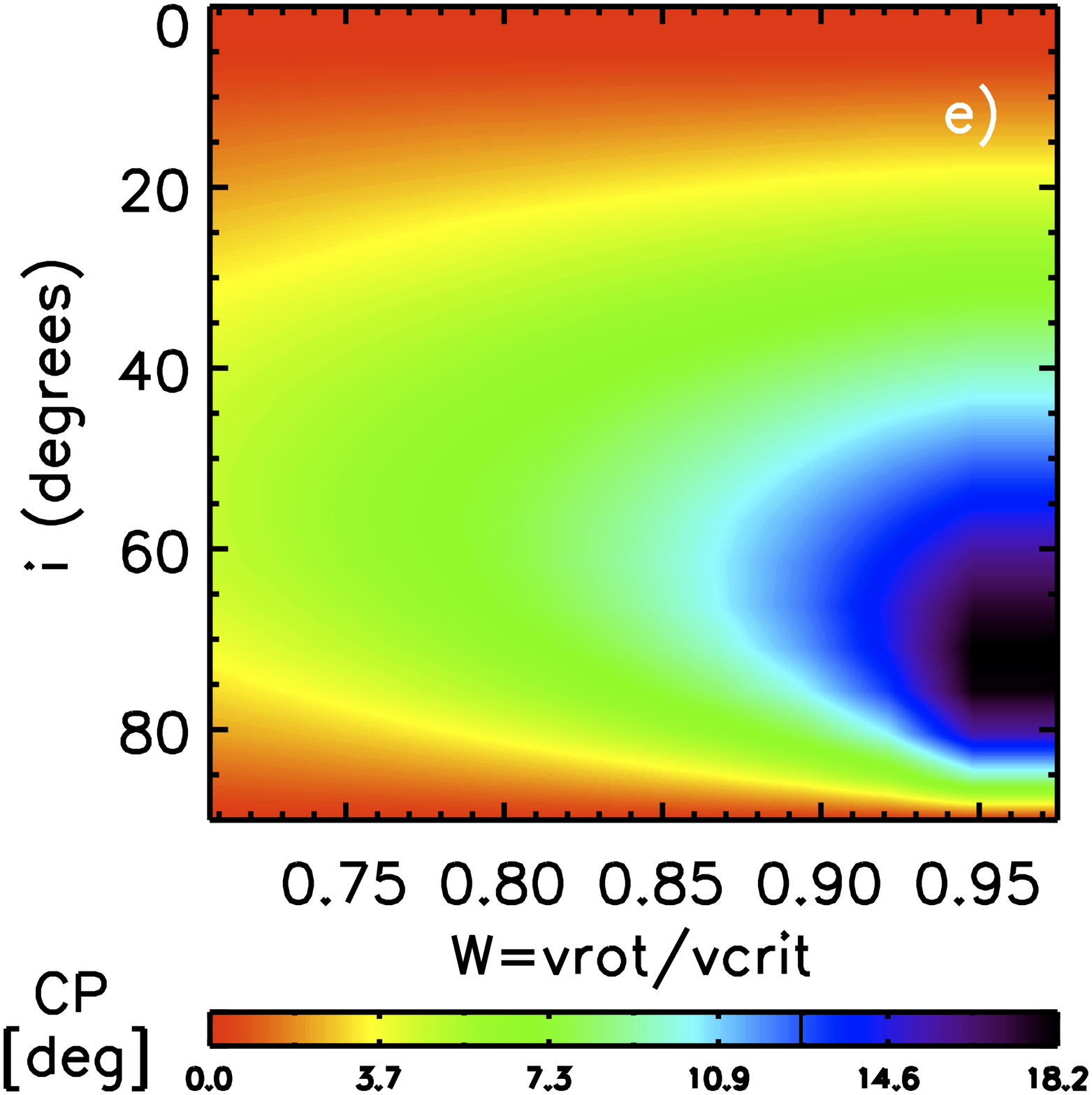}}\\
\resizebox{0.485\hsize}{!}{\includegraphics{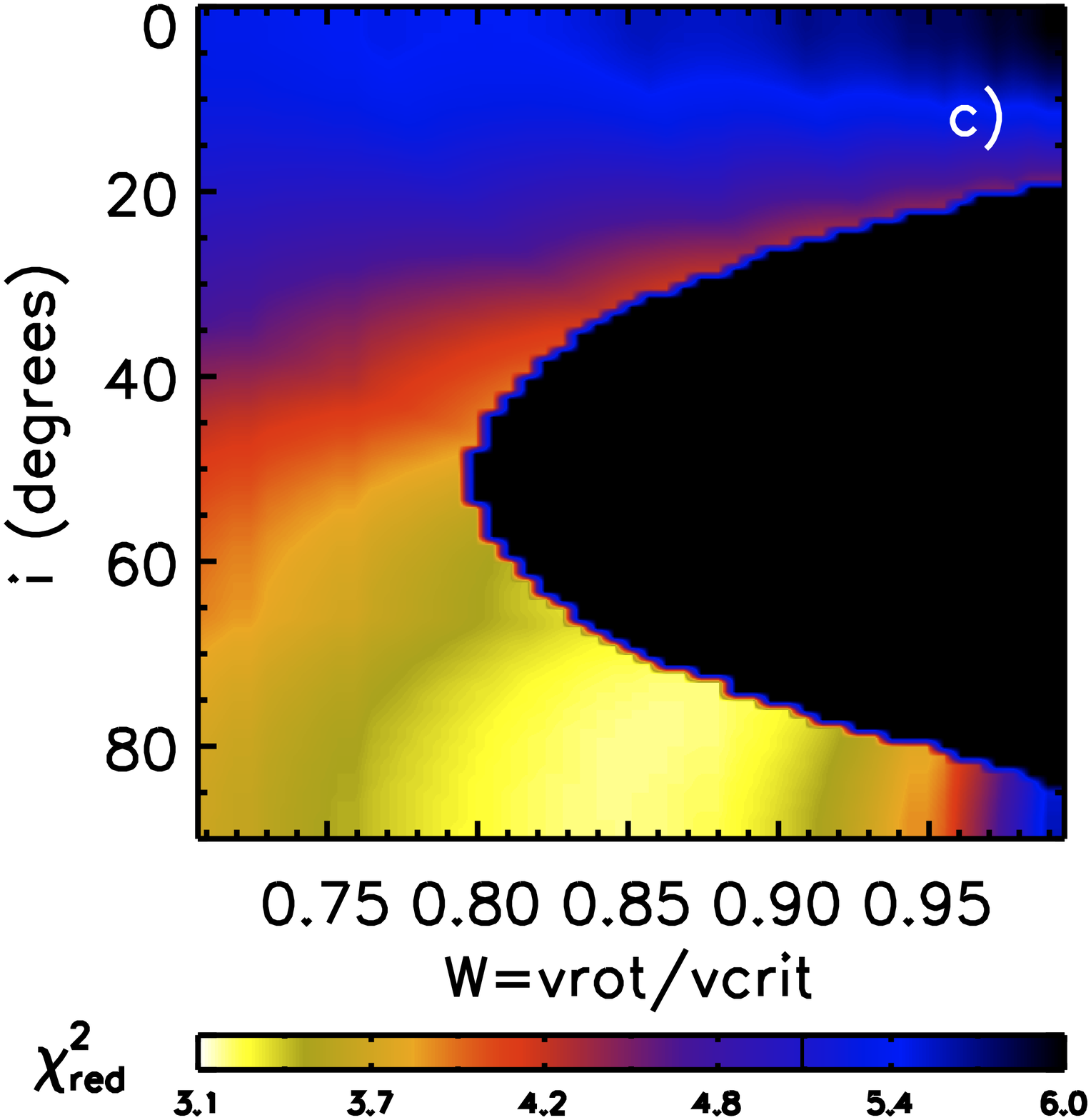}}
\resizebox{0.485\hsize}{!}{\includegraphics{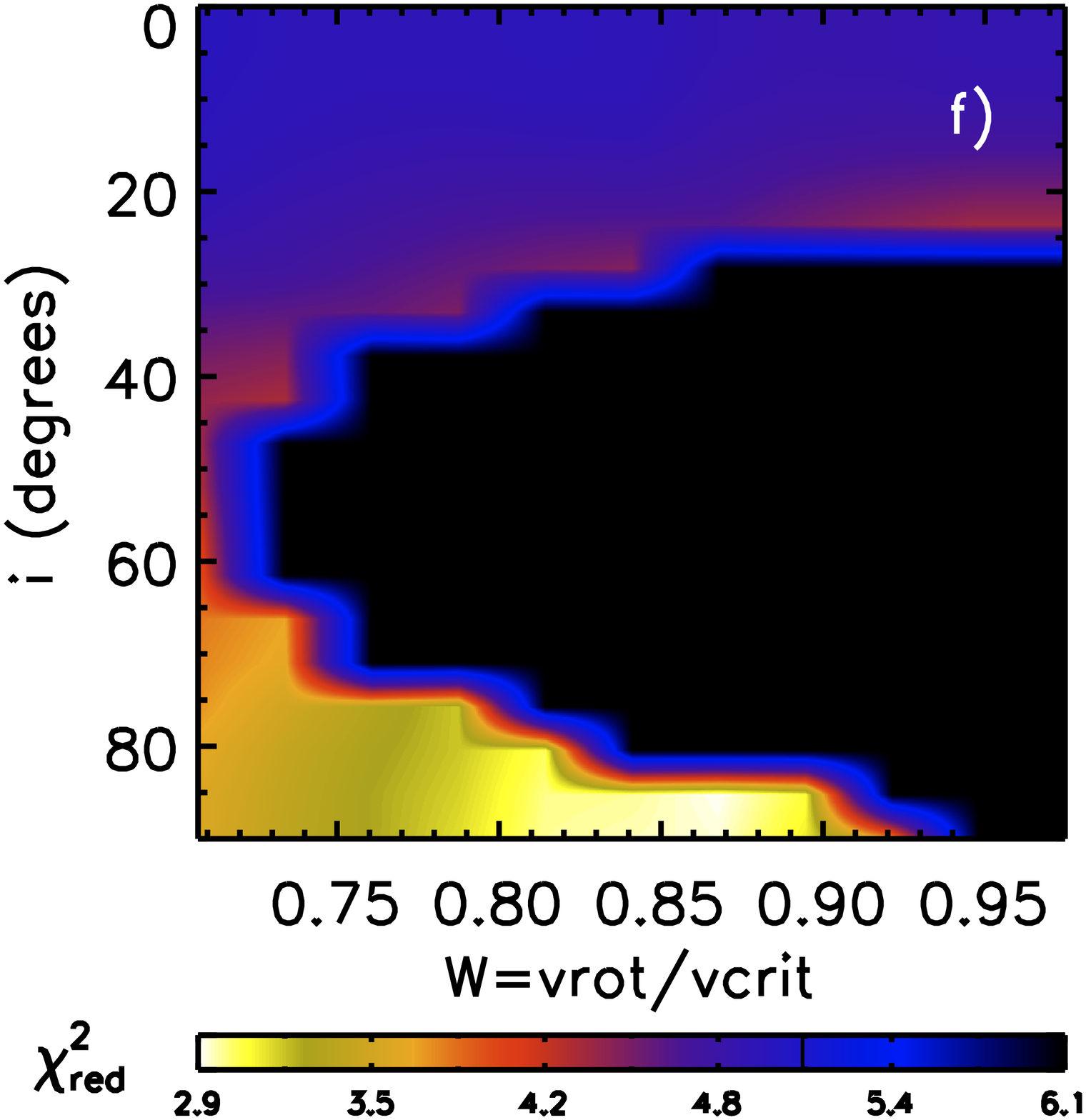}}
\caption{\label{figchi}Left column: From top to bottom, color-coded plots of the reduced $\chi^2$ values of the fit of the VINCI visibilities (a), closure phase (b), and reduced $\chi^2$ values rejecting models with $\mathrm{CP} \geq 5\degr$ (c), as a function of $W$ and $i$, for prolate models with $\mathrm{PA}=130 \degr$. Right column: Idem, for oblate models with $\mathrm{PA}=220 \degr$(d$-$f).}
\end{figure}

Since the spherical CMFGEN model of \citetalias{hillier01} does not fit the geometry of the $K$-band continuum \citepalias{vb03,weigelt07}, we first attempted prolate wind models (Figure \ref{figlatiden}) to reproduce the suggested elongation of the $K$-band continuum observed with VINCI in 2003 January--February. We find the model parameters to be degenerate, as models with $\mathrm{PA}=108\degr$--$142\degr$, $W =$ 0.80--0.99, and $i=30\degr$--$90\degr$ fit the VINCI data equally well. Figure \ref{figchi}a presents a color-coded plot of the reduced $\chi^2$ values of the fit to the VINCI data, with the whitish regions corresponding to the best-fit models. In general, models with relatively large $W$ require lower $i$, while those with relatively lower $W$ need larger $i$ to fit the data.

Additional constraints to $W$ and $i$ can be obtained by analyzing the closure phase, which measures the asymmetry of the brightness distribution.  \citetalias{weigelt07} measured CP$=0\degr \pm 3\degr$ in the K-band continuum in 2005 Feb, and a color-coded plot of the model CPs, computed using the same telescope array configuration used by \citetalias{weigelt07}, is shown in Figure \ref{figchi}b. We find that prolate models with high $W$ (0.82 to 0.99) and intermediate inclination angles ($i\simeq30\degr-60\degr$) yield a K-band image which is noticeably not point-symmetric (Figure \ref{figlatiden}h), and thus are ruled out since they have much larger CPs ($7\degr$ to $19\degr$) than observed. To fit simultaneously the VLTI/VINCI visibilities and the VLTI/AMBER CPs, $W=0.77$--0.92 and $i=60\degr-90\degr$ are needed (Figure \ref{figchi}c). Strikingly, this inclination is significantly higher than that of the Homunculus ($i_\mathrm{Hom}=41\degr$; \citealt{smith06}), which has been assumed so far to align with the {\it current} rotation axis of  $\eta_\mathrm{A}$.  We find that, based on the available interferometric data, if  $\eta_\mathrm{A}$ has a prolate wind, its rotation axis is not necessarily aligned with the Homunculus polar axis.

The image asymmetries from models with $i=30\degr$ to $60\degr$ (red spot in Figure \ref{figlatiden}h) arise from the latitude-dependent K-band photosphere associated with the wind density variations. Rays from the NW hemisphere cross the low-density equator, and so accumulate a lower optical depth than those from the SE hemisphere at the same impact parameter. The deeper penetration of NW vs. SE rays thus leads, for the nearly spherical source function  which is larger for smaller radii, to a stronger emerging radiation field, and thus a higher surface brightness in the NW hemisphere.

\begin{figure*}
\centering
\resizebox{0.245\hsize}{!}{\includegraphics{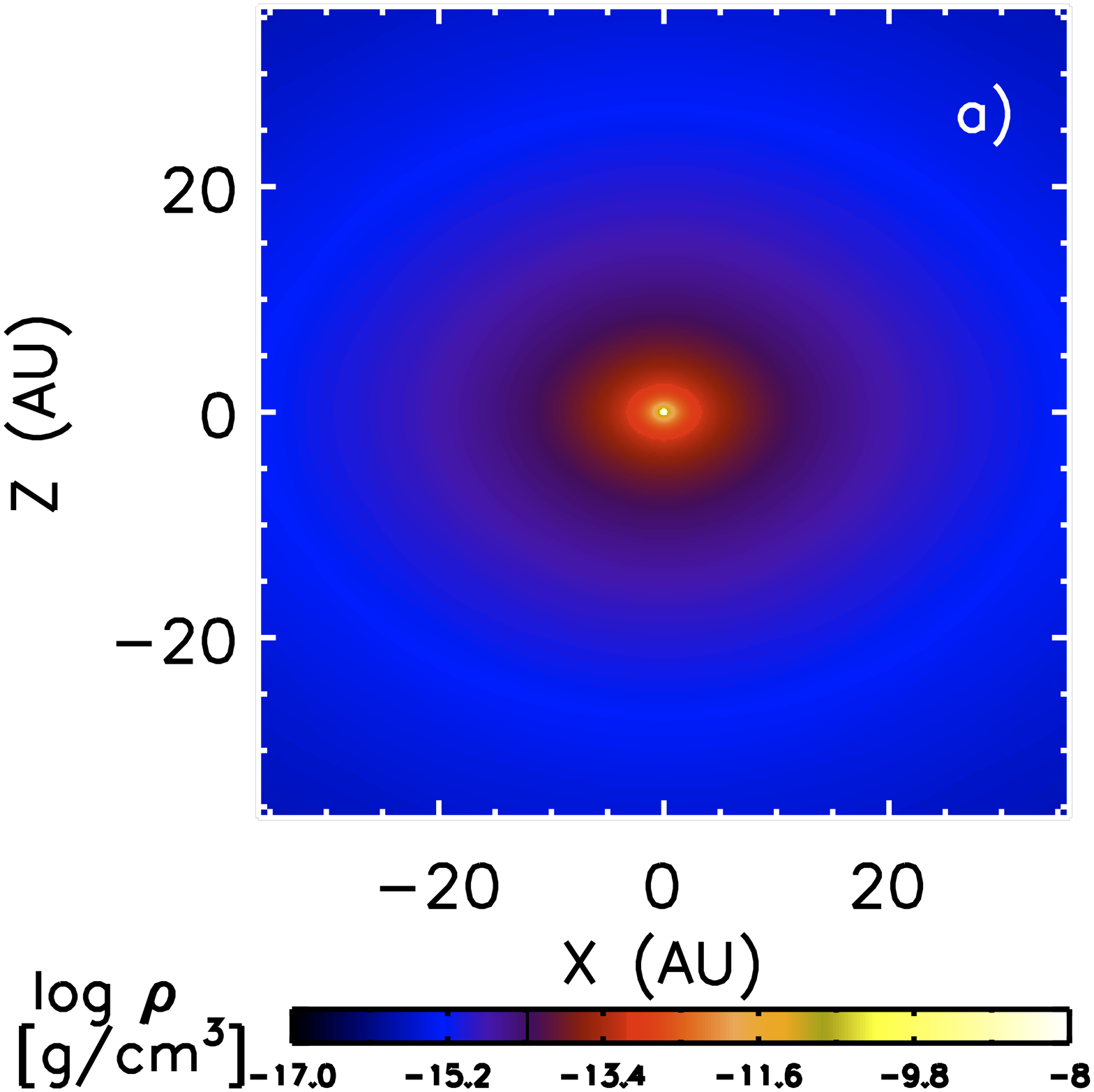}}
\resizebox{0.245\hsize}{!}{\includegraphics{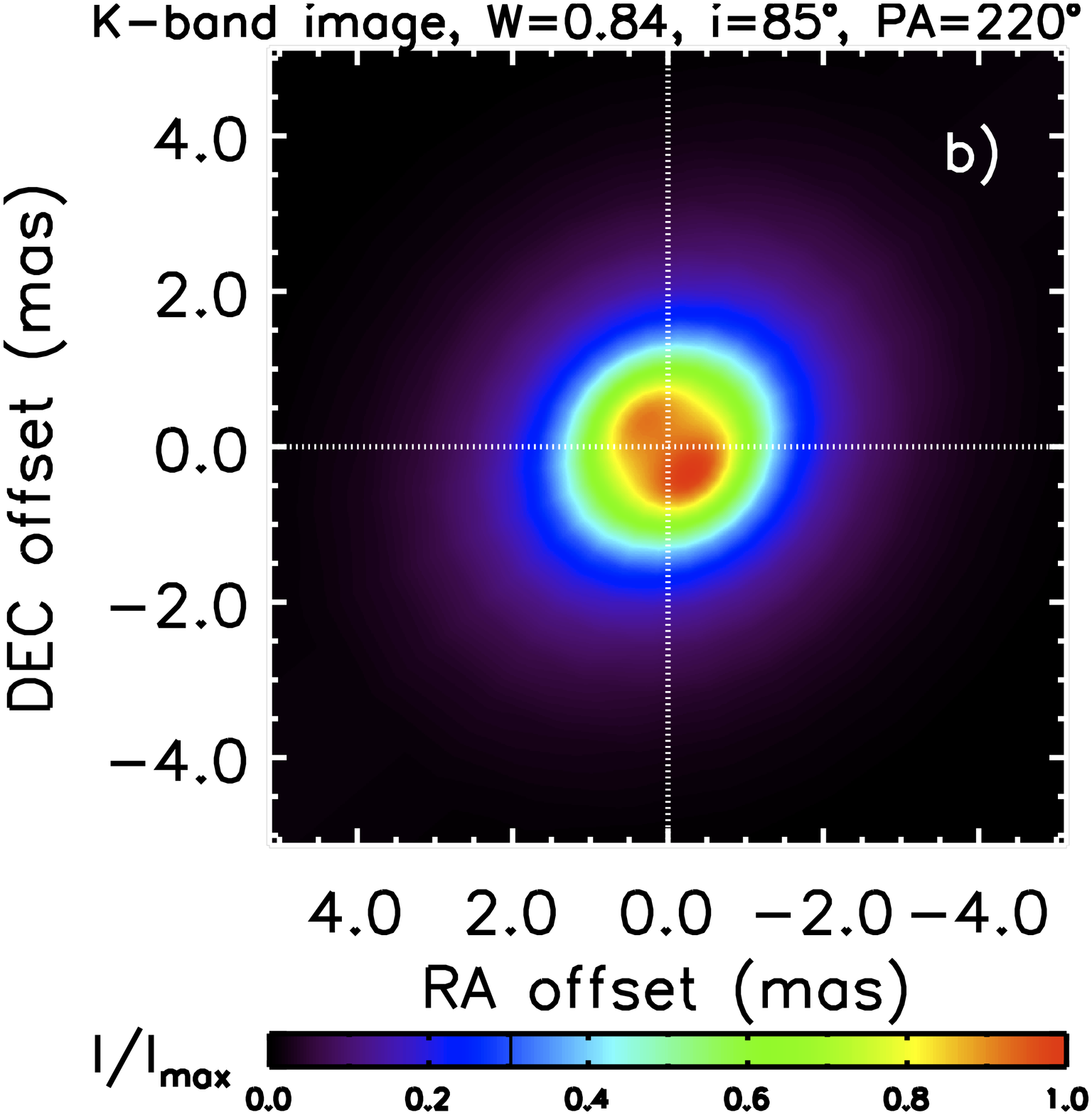}}
\resizebox{0.245\hsize}{!}{\includegraphics{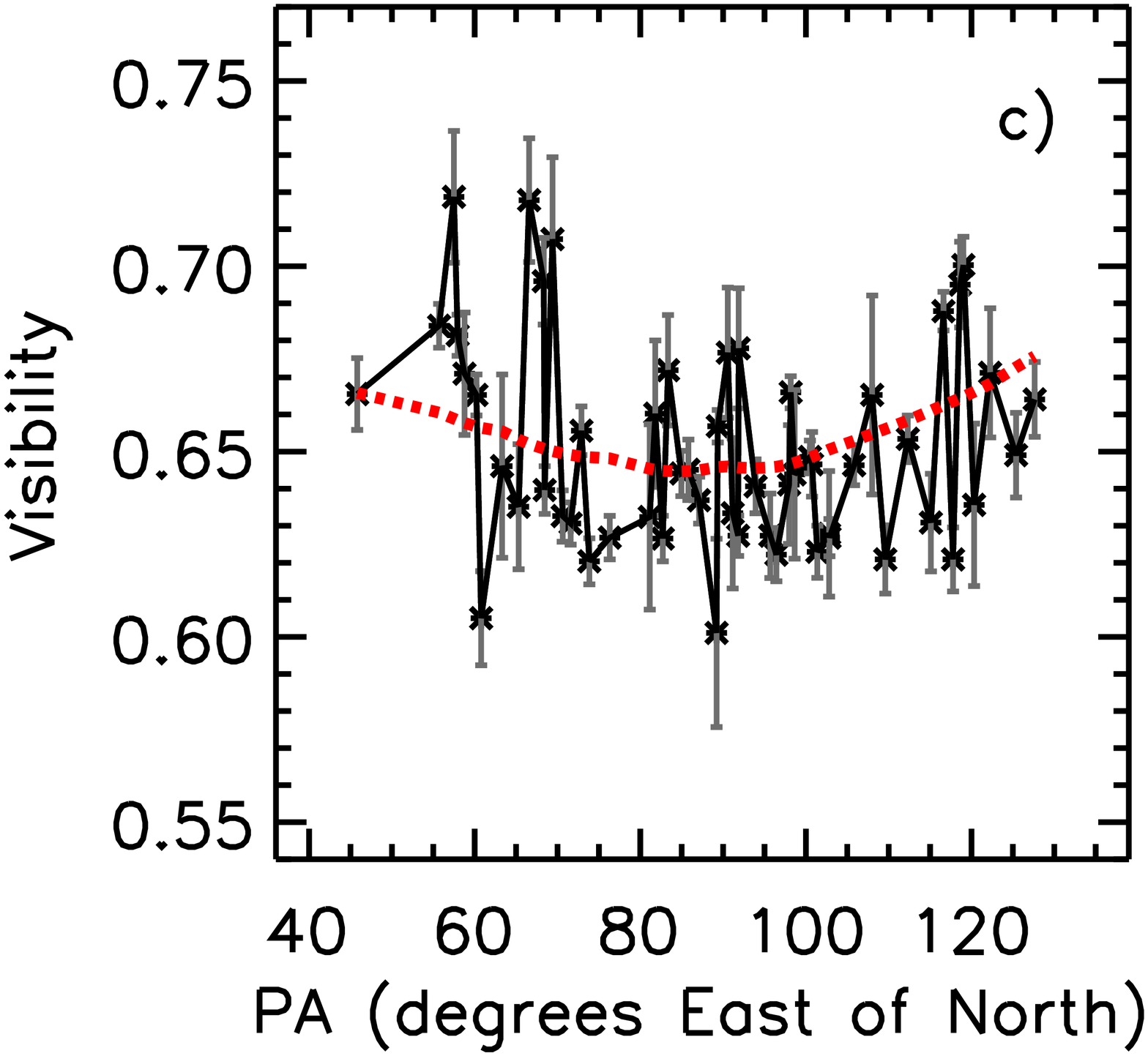}}\\
\caption{\label{figlatiden2} Similar to Figure \ref{figlatiden}, but for a best-fit oblate wind model with  $W = 0.84$, $i=85\degr$, $\mathrm{PA}=220\degr$.}
\end{figure*}

One may ask if a prolate wind model is unique in being able to fit the interferometric data. Therefore, we also analyze the visibility as a function of telescope PA and CP for oblate wind models. We find that the oblate model fits the visibilities as well as the prolate model, provided that $\mathrm{PA}=210\degr-230 \degr$ (Figure \ref{figlatiden2}), i.e., the stellar rotation axis projected on the plane of the sky is rotated by 80\degr--100$\degr$ from the Homunculus symmetry axis. Oblate models with  $W=0.73$--0.90 and $i=80-90\degr$ are able to fit both the observed visibilities and CP (Figure \ref{figchi}d--f). Therefore, for oblate wind models, the rotation axis of $\eta_\mathrm{A}$ is not aligned with the Homunculus polar axis.

\section{Effects of the wind-wind collision zone created by the companion} \label{cavity}

Since Eta Car is believed to be a colliding-wind binary system, we also investigate the effect of the low-density cavity and dense wind-wind interaction region created by the wind of $\eta_\mathrm{B}$ in the wind of $\eta_\mathrm{A}$.

\subsection{Radiative transfer modeling} \label{radtranscav}

Our models are based on the spherically symmetric models of $\eta_\mathrm{A}$ \citepalias{hillier01,hillier06}, with the same stellar parameters as in Section \ref{radtranslat}, but using the 2-D code of \citetalias{bh05} to create a low-density cavity and dense interaction-region walls. Further details are given in Groh et al. 2010 (in preparation), and here we outline only the main characteristics of our implementation.

We approximate the cavity as a conical surface with half-opening angle $\alpha$ and interior density 0.0016 times lower than that of the spherical wind model of $\eta_\mathrm{A}$. We include cone walls of angular thickness $\delta\alpha$ and, assuming mass conservation, a density contrast in the wall of $f_\alpha = [1-\cos(\alpha)]/[\sin(\alpha)\delta \alpha]$ \citep{gull09} times higher than the wind density of the spherical model of $\eta_\mathrm{A}$ at a given radius. The conical shape is justified since the interferometric observations were taken at orbital phases sufficiently before periastron when such a cavity has an approximately 2-D axisymmetric conical form \citep{okazaki08}. Based on the expected location of the cone apex during these phases \citep{canto96,okazaki08}, we place the cavity at a distance $d_{\mathrm{apex}}$ from the primary star. We assume that the material inside the cavity and along the walls has the same ionization structure as the wind of $\eta_\mathrm{A}$, implying  that the shocked wind of the primary star cools radiatively.

\subsection{Results} \label{rescav}

The amount of influence of the wind-wind interaction on the observables depends on two factors: how close the cavity gets to the K-band emitting region of $\eta_\mathrm{A}$ \citep[``bore hole effect",][]{madura10}, and how much free-free radiation is emitted by the dense walls of the shock cone (``wall effect"). We find that the interferometric observables in the K-band are insensitive to the bore hole effect if $d_{\mathrm{apex}} \gtrsim 8~\mathrm{AU}$. This is because most of the observed K-band emission comes from a region with a characteristic 50\% encircled-energy radius of 4.8~AU \citepalias{weigelt07}, and a halo with significant emission exists up to $\sim 8$ AU. For the standard assumed orbital parameters of Eta Car (see, e.g., \citealt{okazaki08}), $d_{\mathrm{apex}} \gtrsim 8~\mathrm{AU}$ during most of the orbit ($0.055 < \phi < 0.945$), which includes the epochs when the VINCI ($\phi_\mathrm{vb03}=0.92$--0.93) and AMBER  ($\phi_\mathrm{w07}=0.27$--0.30) observations  were taken. The VINCI observations were obtained when $d_{\mathrm{apex}} \simeq 10~\mathrm{AU}$, thus a bore hole effect alone is not able to explain the VINCI observations. 

\begin{figure*}[!t]
\centering

\resizebox{0.24\hsize}{!}{\includegraphics{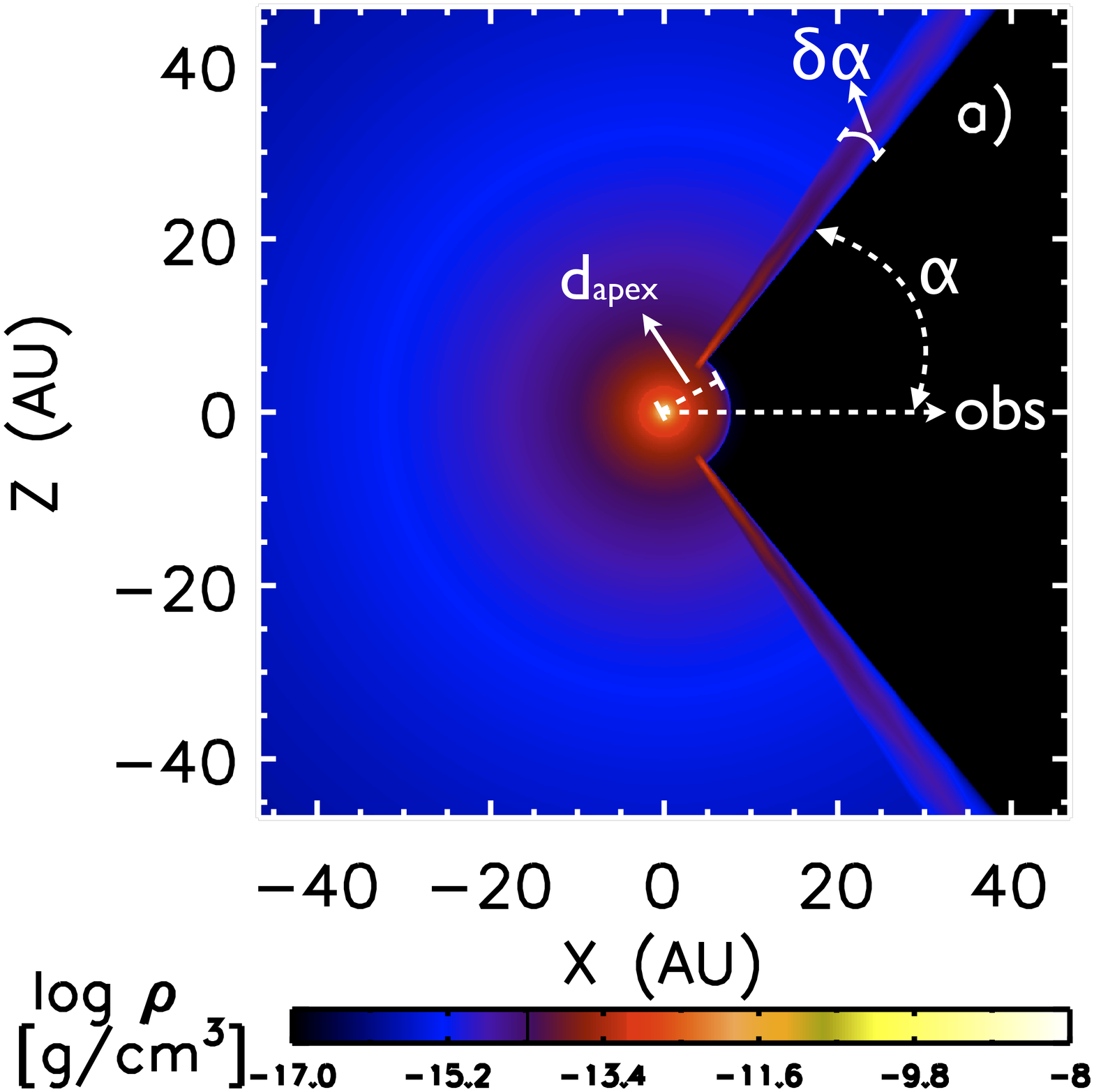}}
\resizebox{0.24\hsize}{!}{\includegraphics{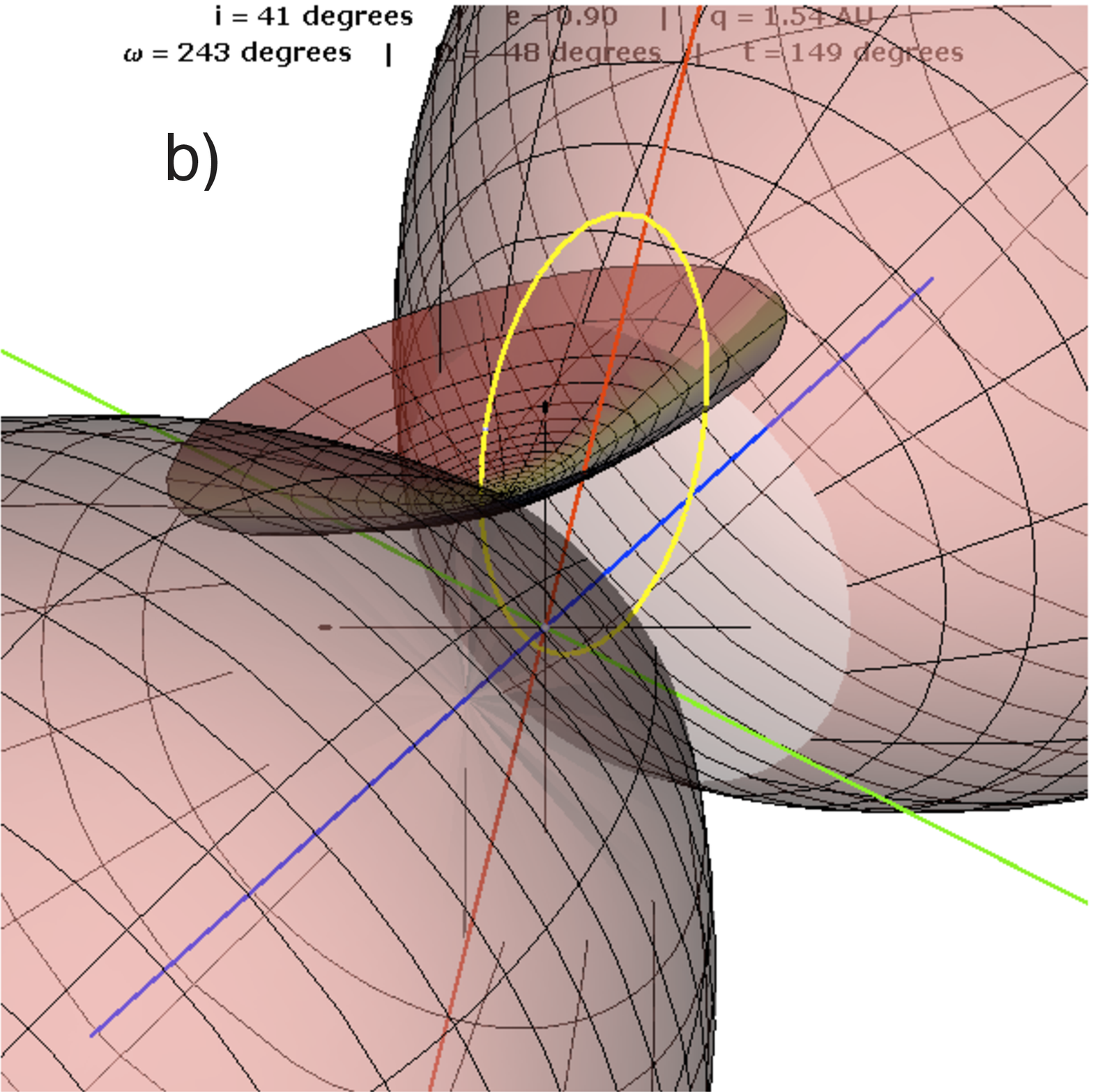}}
\resizebox{0.24\hsize}{!}{\includegraphics{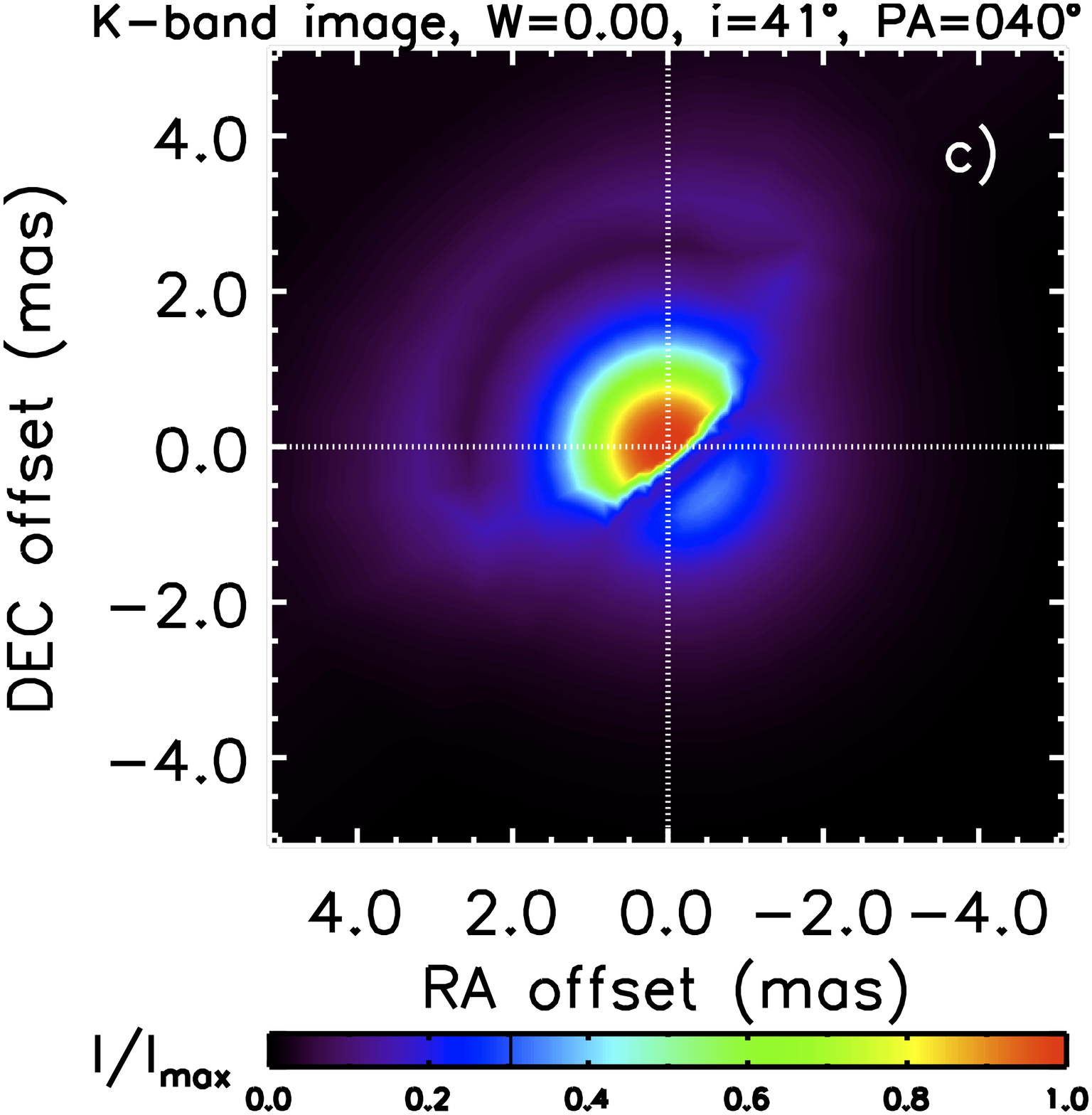}}
\resizebox{0.24\hsize}{!}{\includegraphics{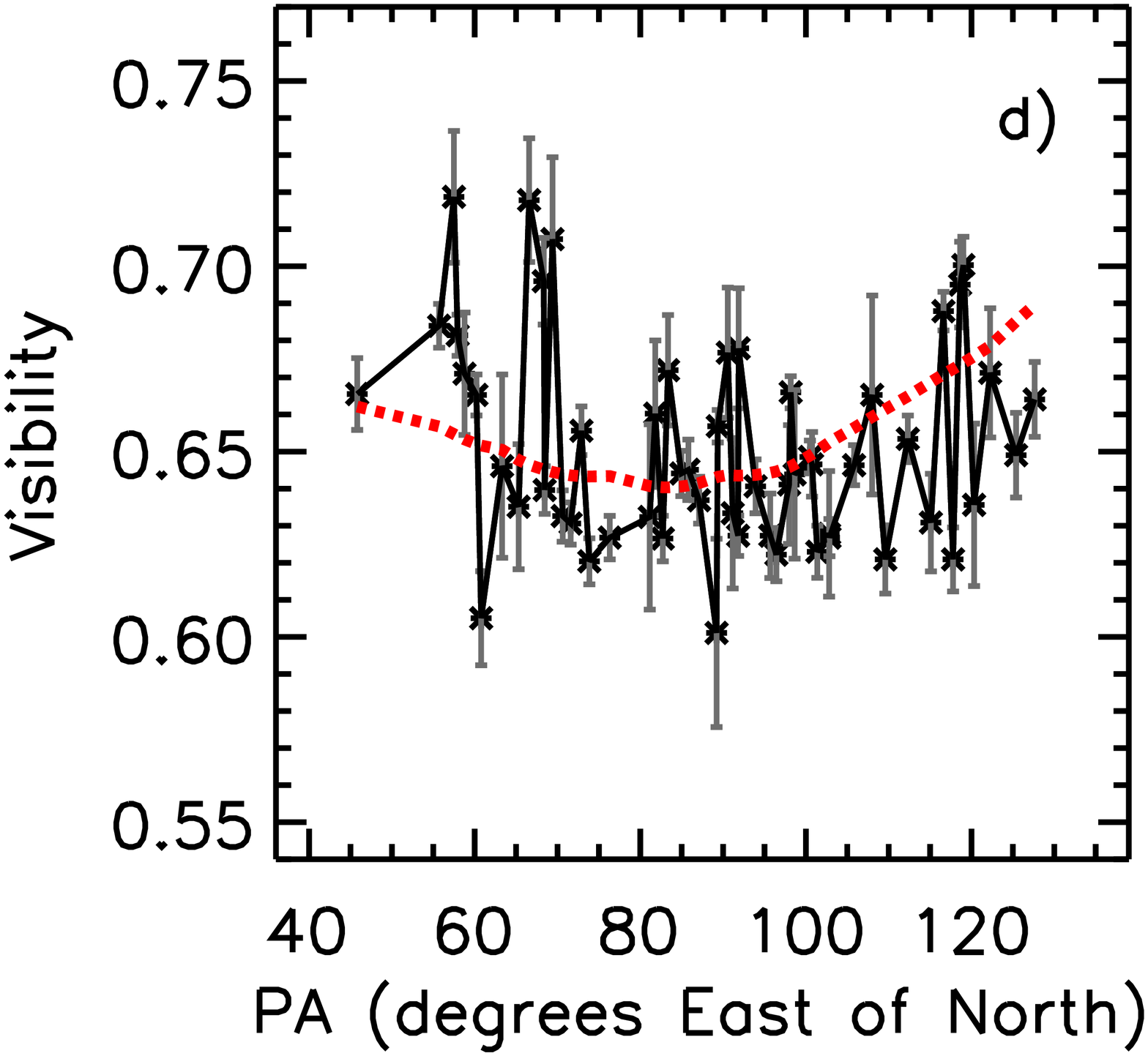}}
\caption{\label{figcav} {\it Panels a,c,d:} Similar to Figure \ref{figlatiden}, but for the $\eta_\mathrm{A}$ model with a spherical wind including a cavity and compressed walls created by the wind of $\eta_\mathrm{B}$. The model is appropriate for the VINCI observations ($\phi=0.93$) and assumes  $i=41\degr$, $\mathrm{PA}=40\degr$, $d_{\mathrm{apex}}=10~\mathrm{AU}$,  $\alpha=54^{\circ}$, and $\delta\alpha=3\degr$. {\it Panel b:} Sketch of the binary orbit (yellow) and the shock cone orientation at $\phi=0.93$ relative to the Homunculus (not to scale), assuming a counterclockwise motion of $\eta_\mathrm{B}$ on the sky. The orbital plane is assumed to be in the skirt plane with the orbital axis (blue) aligned to the Homunculus axis of symmetry. North is up and East is to the left.}
\end{figure*}

According to our models, the main effect from the wind-wind collision zone at the orbital phases analyzed is extended free-free emission from the compressed walls. Such an effect may influence the geometry of the K-band emitting region even at orbital phases far from periastron, depending on the observer's location, geometry of the cavity,  $i$, $\alpha$, and $f_\alpha$. A model with $\alpha=54^{\circ}$, $f_\alpha=9.7$ (i. e., $\delta\alpha=3\degr$; \citealt{gull09}), and $i=41\degr$  shows significant wall emission and is able to produce a significant elongation of the K-band continuum (Figure \ref{figcav}). To explain the VINCI observations, the symmetry axis of the cavity has to be oriented along PA$\simeq35\degr-45\degr$ (i.e., SW-NE axis; Fig. \ref{figcav}), which is roughly consistent with that expected from a longitude of periastron of $\omega=243\degr$ \citep[e.g,][]{okazaki08,parkin09,gull09,groh10} and a counterclockwise motion of $\eta_\mathrm{B}$ on the sky  (Figure \ref{figcav}b).
The $\mdot$ of $\eta_\mathrm{A}$ needs to be slightly reduced to $ 8\times 10^{-4}~\msunyr$ to compensate for the extra extension in the K-band emitting region caused by the wall effect. Comparing Figures \ref{figlatiden} and \ref{figcav}, the cavity and walls have an effect on the available interferometric observables that is as large as that due to a latitude-dependent wind caused by rapid rotation, although the morphology of the K-band images is noticeably different.

The cavity model assumed here is very idealized, since it does not take into account the distortion of the shock cone \citep{okazaki08}, or the instabilities thought to arise in the wind-wind collision zone \citep{parkin09}. Nevertheless, our conclusion is that, using a 2-D model, noticeable elongation of the K-band region can occur if the apex of the cone penetrates a significant distance into the K-band emitting region, that cone is open toward the observer, and/or the compressed walls are dense enough and emit free-free radiation. Three-dimensional radiative transfer models, when available, would be desirable to test the conclusions found here.

\section{Discussion} \label{disc}

$\eta_\mathrm{A}$ is routinely referred to as the prototype of a massive star with a fast, dense polar wind created by rapid stellar rotation, even though the system is believed to contain a massive companion, $\eta_\mathrm{B}$. How much does $\eta_\mathrm{B}$ affect the K-band emitting region of $\eta_\mathrm{A}$?  We show that, assuming the standard orbital and wind parameters of Eta Car, even if $\eta_\mathrm{A}$ has a spherical wind, its inner density structure can be sufficiently disturbed by $\eta_\mathrm{B}$, mimicking the effects of a prolate/oblate latitude-dependent wind in the available interferometric observables in the K-band continuum. Therefore, fast rotation may not be the only explanation for the interferometric observations.

If we ignore the presence of the companion, our 2-D modeling shows that both single star prolate and oblate wind models are able to explain the elongation of the K-band emitting region of $\eta_\mathrm{A}$ and the CP measurements. While a prolate wind is thought to arise in a gravity-darkened, fast-rotating star with a radiative envelope \citep{owocki96,owocki98}, an oblate wind can be produced by a fast-rotating star when gravity-darkening is not important \citep{bc93,owocki94,owocki96,owocki98,ignace96}, for instance if the envelope is predominantly convective. Interestingly, convection can be present in the envelope of stars close to the Eddington limit \citep{langer97,maeder08}, such as  $\eta_\mathrm{A}$, although it is unclear whether convection is efficient enough to transport a significant fraction of the flux. 

We find that moderately fast rotation and high inclination angles are required to fit simultaneously the VINCI and AMBER data, with $W=0.77$--0.92 and $i=60\degr-90\degr$ for the best prolate models, and $W=0.73$--0.90 and $i=80\degr-90\degr$ for the best oblate models. The possible tilting between the current rotation axis of $\eta_\mathrm{A}$ and the Homunculus axis is striking. If the Giant Eruption, which created not only the Homunculus nebula, but also a blast wave with much higher velocities \citep{smith08b}, happened in $\eta_\mathrm{A}$, that would indicate that either the angular momentum loss was not axisymmetric, that strong dynamical interactions between $\eta_\mathrm{A}$ and $\eta_\mathrm{B}$ (or a third member) occurred in the mean time, or both.

Alternatively, a speculative scenario that permits $i=i_\mathrm{Hom}=41\degr$ is one that allows for time variability in $W$, which means a different latitude dependency of the prolate wind at the different epochs of the observations. This would require $W=0.93-0.99$ during the VINCI observations in 2003 January--February ($\phi_\mathrm{vb03}=0.92$--0.93), and a significantly decreased $W=0.75-0.82$ during the AMBER observations in 2005 Feb ($\phi_\mathrm{w07}=0.27$--0.30 ). A possible explanation for this variability could be a change in the stellar radius due to S-Dor type variability, which would cause $\vrot$ and $W$ to vary, as has been seen for other LBVs such as AG Car \citep{ghd06,ghd09} and HR Car \citep{gdh09}
\footnote{A variability in $\vrot$, with a gradual increase before the spectroscopic event, has been suggested before (\citetalias{smith03}; \citealt{davidson99}), but in a different context (shell ejections). }.
 Regardless of such a speculative scenario, based on the available interferometric data alone, we are unable to determine whether $\eta_\mathrm{A}$ possesses an oblate or prolate wind, and whether or not the rotation axis is aligned with the Homunculus symmetry axis (both $i$ and PA). The fact that the wind-wind collision zone can mimic the effects of a latitude-dependent wind further complicates matters. Thus, more interferometric observations are needed to resolve these issues.

Much of the support for the presence of a fast, dense polar outflow comes from spectroscopic observations of the H$\alpha$ absorption profile at different locations in the Homunculus (\citetalias{smith03}; \citealt{hillier92}). Assuming that the rotation axis of $\eta_\mathrm{A}$ is aligned with the Homunculus axis, \citetalias{smith03} found that the absorption profile of H$\alpha$ extends to velocities as high as $-1200~\kms$ in the polar spectrum, while at mid-latitudes a value of $-600~\kms$ is found. However, as pointed out by \citetalias{smith03}, the latitudinal behavior of the H$\alpha$ absorption can be affected by the presence of $\eta_\mathrm{B}$ if it is able to photoionize the region where H absorption would occur. Determining how much the wind terminal velocity changes as a function of latitude is only possible when the influence of $\eta_\mathrm{B}$ is properly determined. The exact influence of $\eta_\mathrm{B}$ is unknown and depends mainly on the number of its H ionizing photons and the orbital parameters. It would be intriguing to investigate how $\eta_\mathrm{B}$ may affect these H$\alpha$ diagnostics. Detailed multi-dimensional radiative transfer models of spectral lines and other diagnostics are urgently warranted, and will be the subject of future work.

\acknowledgments

We thank Roy van Boekel and Pierre Kervella for making the VINCI data available, and an anonymous referee for comments. JHG is supported by the Max Planck Society, and TIM by a NASA GSRP fellowship.

\end{document}